%% file: main.tex
\documentclass[reprint, superscriptaddress, showkeys,
aps, prl, showkeys,
amsmath,amssymb,
]{revtex4-2}

\usepackage{graphicx}
\graphicspath{{./figures/}}
\usepackage[svgnames]{xcolor}
\usepackage{dcolumn}
\usepackage{multirow}
\usepackage{makecell}
\usepackage{bm}
\usepackage{siunitx}
\DeclareSIUnit\angstrom{\text {Å}}
\DeclareSIUnit\muB{\ensuremath{\mu_{\mathrm{B}}}} 
\usepackage[normalem]{ulem}    
\newcommand\change[1]{{\color{black}#1}}

\usepackage[hidelinks]{hyperref}
\usepackage[mathlines]{lineno}

\begin{document}
	
	
	\title{Quasi-aperiodic grain boundary phases of $\Sigma5$ tilt grain boundaries in refractory metals}
	
	\author{Enze Chen}
	\email{enze@stanford.edu}
	\affiliation{Department of Materials Science and Engineering, Stanford University, Stanford, CA 94305, USA}
	
	\author{Timofey Frolov}
	\email{frolov2@llnl.gov}
	\affiliation{Materials Science Division, Lawrence Livermore National Laboratory, Livermore, CA 94550, USA}

	\date{\today}
	
	\begin{abstract}
		We report new ground-state structures and phase transitions in $\Sigma$5[001] tilt grain boundaries (GBs) in body-centered cubic (BCC) refractory metals Nb, Ta, Mo, and W. 
		$\Sigma$5 tilt GBs have been extensively investigated over the past several decades, 
		with their ground-state structure---composed of kite-shaped structural units---previously thought to be well understood. 
		By performing a rigorous GB structure search that optimizes the number of atoms in the boundary core, we predict new quasi-aperiodic ``split kite" phases analogous to those previously found in GBs in face-centered cubic metals. 
		Our results suggest that complex aperiodic phases of GBs appear to be a general phenomenon, as validated through density functional theory calculations.
		\change{Moreover, the atoms in the split kite phase demonstrate distinct collective diffusion dynamics.}
		Phase-contrast image simulations of split kites show better agreement with experimental observations, offering an alternative explanation for previous microscopy results and motivating future atomically resolved imaging of the GB structure.
	\end{abstract}
	
	\keywords{grain boundary, phases, optimization}
	
	\maketitle

	
	
	\emph{Introduction}: 
	Most engineering materials are composed of many differently oriented crystallites, or grains, separated by interfaces called grain boundaries (GBs). 
	These interfaces have structures and properties distinct from the bulk material and can greatly influence material properties~\cite{sutton_1995}. 
	Among the vast space of possible GBs described by the five macroscopic degrees of freedom, a special subset is the coincidence site lattice boundaries, for which the adjoining crystals form interfaces with well-defined periodic unit cells~\cite{ranganathan_1966}. 
	$\Sigma5$(210)[001] and $\Sigma5$(310)[001] tilt GBs with their small periodic units serve as convenient model systems as representative high-angle, high-energy boundaries for both density functional theory (DFT) calculations and empirical potentials, and have been extensively investigated over the past several decades.

\begin{figure*}[!ht]    
		\includegraphics[width=0.9\linewidth]{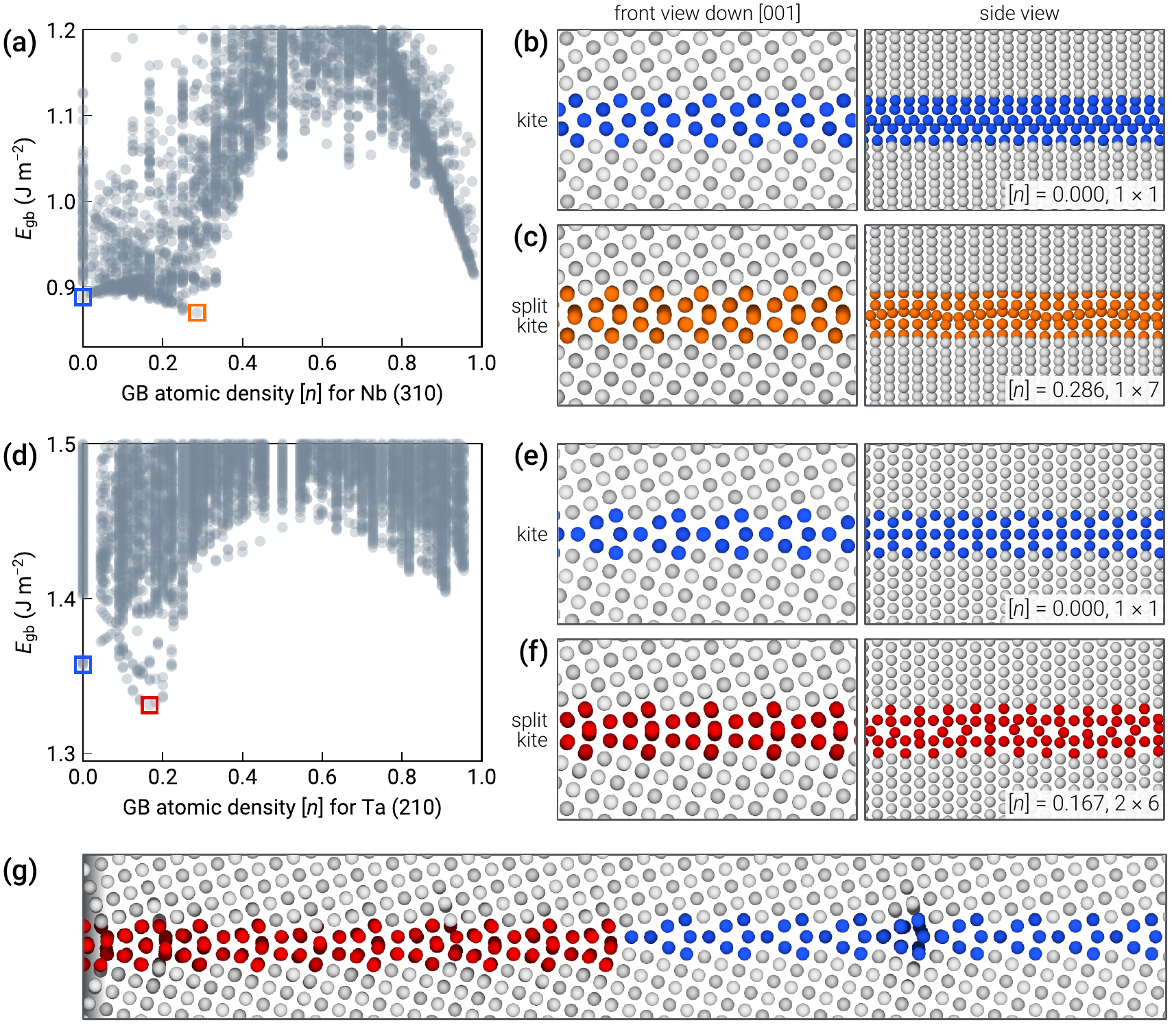}
		\caption{\textbf{Predicted split kite phases in $\Sigma5[001]$ tilt grain boundaries (GBs) in body-centered cubic metals.}
			(a) GRIP structure search results for Nb (310) are plotted as $E_{\mathrm{gb}}$ vs. $[n]$.
			Each gray circle corresponds to a structure generated during the search and notable minima are boxed.
			Two orthogonal views are shown for the minimum-energy 
			(b) kite and
			(c) split kite structures.
			(d) GRIP results for Ta (210), along with the corresponding minimum-energy
			(e) kite and 
			(f) split kite structures. 
			(g) First-order phase transitions are observed with open-surface boundary conditions in finite-temperature molecular dynamics simulations ($T = \SI{1750}{\kelvin}$).
			All GB atoms are colored to highlight their distinct coordination.
		}
		\label{fig:grip_NbTa}
	\end{figure*}

	The structure of the $\Sigma5$ boundaries is considered well understood and composed of ordered kite-shaped structural units shown in \autoref{fig:grip_NbTa}b and \ref{fig:grip_NbTa}e. 
	This kite-shaped structure was experimentally observed by atomic-resolution scanning transmission electron microscopy (STEM) only in face-centered cubic (FCC) Cu~\cite{ding_2024} and body-centered cubic (BCC) Fe~\cite{medlin_2017, zhou_2023}, and inferred through phase contrast imaging in other elemental metals~\cite{shamsuzzoha_1996, cosandey_1990, campbell_1991, campbell_1993, campbell_2000, campbell_1999, bacia_1997, morita_1997}. 
	It was also independently predicted in simulations with empirical potentials~\cite{sutton_1983, wolf_1989}, and density functional theory (DFT) calculations~\cite{ochs_2000a, scheiber_2016, zheng_2020} that used a traditional modeling approach known as the $\gamma$-surface method~\cite{mishin_1998} that joins two misoriented crystals without optimizing the number of atoms in the boundary core.

	More recently, improved simulation methodologies that allowed the number of atoms in the GB to change revealed that most empirical embedded-atom method potentials of FCC metals predict a completely different ground-state structure, which was named ``split kites"~\cite{frolov_2013_gbphase}. 
	This structure required inserting or removing about half of the plane of atoms in the boundary core. 
	Moreover, split kites exist in multiple energy-degenerate states characterized by different arrangements of their structural units without a well-defined periodic unit cell~\cite{zhu_2018}. 
	Due to its large area, the split kite structure in FCC metals has not been investigated with DFT calculations; 
	however, subsequent studies of [110] tilt GBs in BCC metals revealed \change{other} complex aperiodic structures that were confirmed by DFT calculations~\cite{frolov_2018} and subsequently validated by direct observations in BCC Fe~\cite{seki_2023}.
	Complex GB phases have demonstrated impacts on properties such as GB migration under shear~\cite{pemma_2024} and GB diffusivity~\cite{rajeshwari_2020}.

	Motivated by these studies, we revisit $\Sigma5$ tilt GBs in BCC refractory metals represented by Nb, Ta, Mo, and W. 
	We perform GB structure search using a recently developed Grand canonical Interface Predictor (GRIP) code that optimizes the GB core configuration~\cite{chen_2024}. 
	We discover new phases of these $\Sigma5$ tilt boundaries that are analogous to the split kites observed in FCC metals. 
	These complex structures are confirmed using DFT to be the ground states in Ta, Mo, and W, and are at least metastable in the other BCC systems studied.

	\emph{Results}: 
	In this work, we focus on two symmetric tilt $\Sigma5$(310)[001] and $\Sigma5$(210)[001] boundaries with misorientation angles around the common tilt axis of \SI{36}{\degree} and \SI{53}{\degree}, respectively. 
	The energy of the system is modeled using empirical potentials and DFT calculations. 
	Using GRIP, we systematically explore GB structures by uniformly sampling relative translations between the two grains, randomly removing a fraction of atoms from the boundary plane, and dynamically sampling different configurations using empirical potentials~\cite{chen_2024}.
	As described in our previous studies~\cite{frolov_2013_gbphase}, we track the GB atomic density, $[n]$, by calculating
	
	\begin{equation}
		[n] = \frac{N_{\mathrm{gb}}\ \text{mod}\ N_{\mathrm{plane}}}{N_{\mathrm{plane}}} \in [0, 1),
		\label{eq:n}
	\end{equation}
	
	\noindent where $N_{\mathrm{gb}}$ is the total number of atoms in a region containing the GB (in our case, the entire simulation cell), and $N_{\mathrm{plane}}$ is the number of atoms in the perfect crystal plane parallel to the boundary. 
	GRIP also explores larger GB area reconstructions by systematically tiling the smallest periodic cell. 
	For each GB degree of freedom described above, dynamic sampling is performed by running molecular dynamics (MD) simulations at different temperatures to rearrange the atoms and find the optimal GB configuration. 
	During the final stage of the optimization, the structures are relaxed at \SI{0}{\kelvin} by a conjugate gradient algorithm to calculate the GB energy as
	
	\begin{equation}
		E_{\mathrm{gb}} = \frac{E^{\mathrm{gb}}_{\mathrm{total}} - N^{\mathrm{gb}}_{\mathrm{total}} E^{\mathrm{bulk}}_{\mathrm{coh}}}{A^{\mathrm{gb}}_{\mathrm{plane}}},
		\label{eq:Egb}
	\end{equation}
	
	\noindent where $E^{\mathrm{gb}}_{\mathrm{total}}$ and $N^{\mathrm{gb}}_{\mathrm{total}}$ are the total energy and number, respectively, of atoms in the GB region, $E^{\mathrm{bulk}}_{\mathrm{coh}}$ is the cohesive energy per atom in the bulk structure, and $A^{\mathrm{gb}}_{\mathrm{plane}}$ is the area of the GB plane.

	\begin{figure*}[!ht]    
		\includegraphics[width=0.9\linewidth]{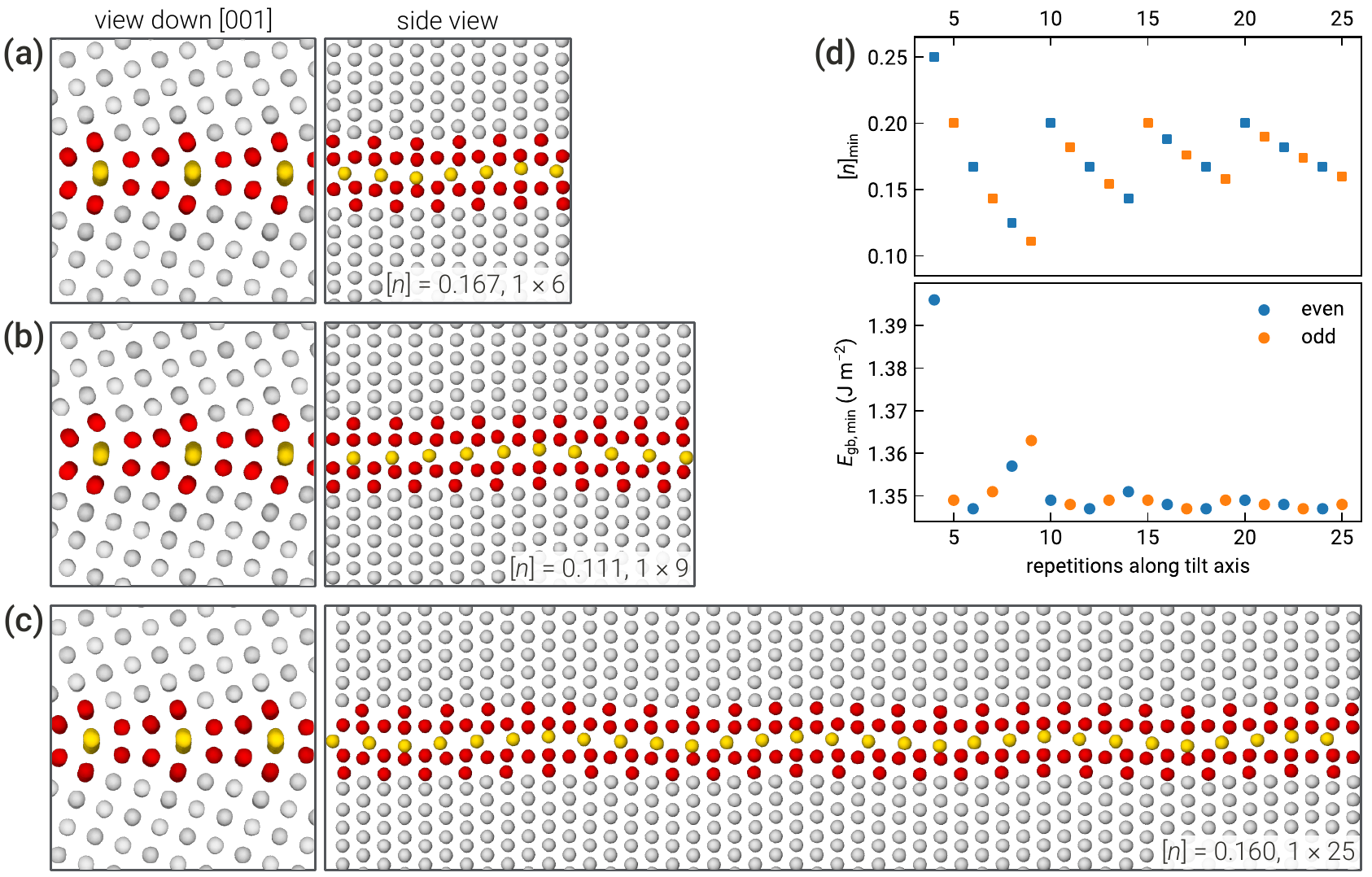}
		\caption{\textbf{The split kite phase is quasi-aperiodic along the tilt axis.}
			Representative results are shown for $\Sigma5(210)[001]$ in Ta for 
			(a) 1~$\times$~6, 
			(b) 1~$\times$~9, and 
			(c) 1$\times$~25 reconstructions of the smallest periodic GB cell. 
			The middle atoms are colored yellow to highlight their distinct wave-like pattern.
			(d) The minimum GB energy and corresponding $[n]$ are plotted against the number of repetitions along [001].
		}
		\label{fig:aperiodic}
	\end{figure*}

	Representative results of GRIP searches for two $\Sigma5$ boundaries in Nb and Ta are shown in \autoref{fig:grip_NbTa}. 
	Each gray circle corresponds to a structure generated during the GRIP search. 
	The \SI{0}{\kelvin} GB energy of each structure is plotted versus the fraction of GB atoms $[n]$. 
	The lowest-energy states at $[n]=0$ indicated by blue squares correspond to the well-known regular kite GB phases of these boundaries shown in panels (b) and (e), respectively. 
	$[n]=0$ indicates that regular kite structures can be obtained simply by joining two misoriented half-crystals without inserting or removing atoms. 
	This is, however, not the lowest-energy configuration of either boundary. 
	The ground-state structures found in this work correspond to $[n]=0.286$ and $[n]=0.167$ for $\Sigma5$(310) in Nb and $\Sigma5$(210) in Ta, respectively, which means that 2/7 and 1/6 of the (310) and (210) crystal planes, respectively, have to be inserted to transform the regular kite structures into the ground states. 
	The ground-state structures for Nb (310) and Ta (210) are shown in panels (c) and (f), respectively. 
	The front view of the boundary structure still has the characteristic shape of kites, but in the atomic column corresponding to the tip of the kite, the atoms have different positions resulting in splitting. 
	Following the naming convention established by previous studies in FCC metals that found similar structural features~\cite{frolov_2013_gbphase, zhu_2018}, we call the new ground states ``split kites." 
	The side views shown in panels (c) and (f) reveal complex arrangement of the atoms in the BCC split kite structures. 
	In the boundary core, they are not aligned with the (002) atomic planes like in the regular kite structure, but occupy positions in between.
	The complex core structures in all split kite phases require larger reconstructions and insertion of additional atoms to be properly optimized.

	\begin{figure}[!ht]   
		\includegraphics[width=\linewidth]{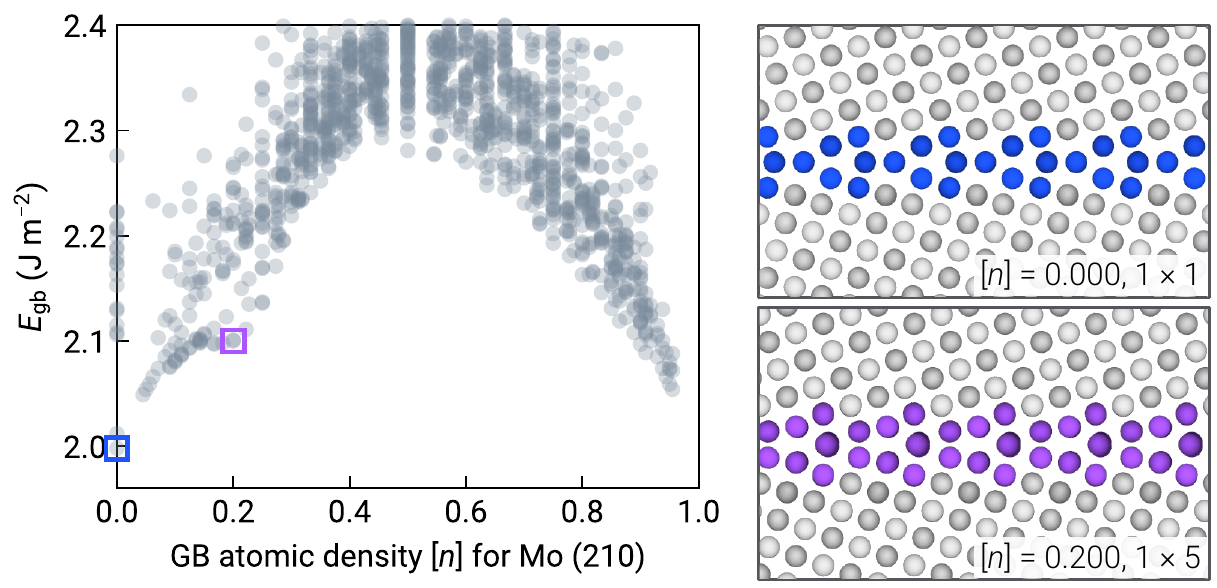}
		\caption{\textbf{The split kite phase in $\Sigma5(210)$ in Mo is predicted to be metastable.}
			In this system, the split kite phase ($[n]=0.2$) has higher energy than regular kites ($[n]=0.0$) and is a local minimum with respect to GB atomic density.
			GB atoms are colored for emphasis.
		}
		\label{fig:grip_Mo}
	\end{figure}

	\begin{figure*}[!ht]    
		\includegraphics[width=0.9\linewidth]{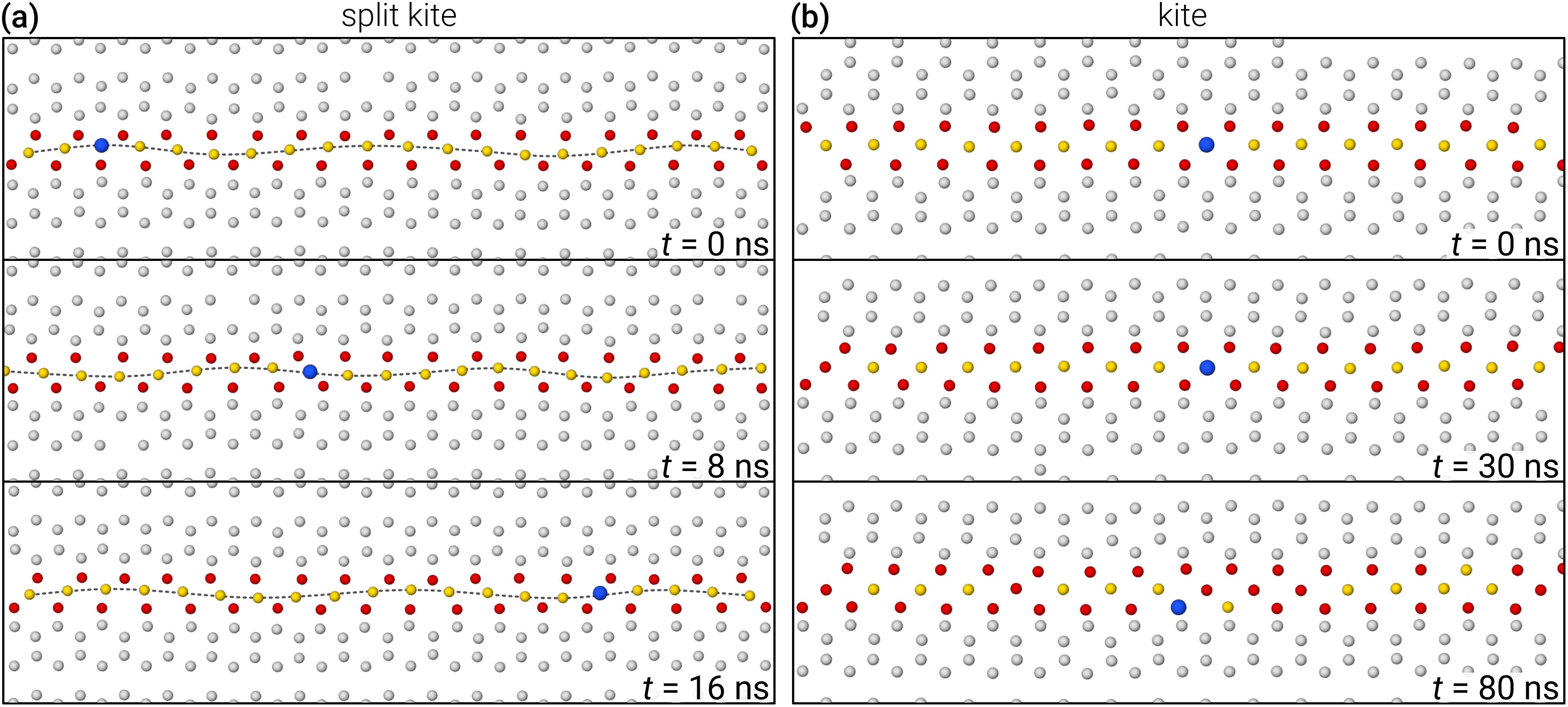}
		\caption{\change{\textbf{GB atoms in split kites exhibit collective dynamics resulting in faster diffusion compared to uncorrelated hops in kites.}
				(a) The wave-like pattern in the tips of split kites (dashed curve as a guide) persists at \SI{1500}{\kelvin} and facilitates accelerated GB diffusion through collective wave motion of the atoms.
				For clarity of visualization, atom positions are averaged over thermal vibrations, only a single wave (tip column) is shown, and a single atom is marked blue to visualize the displacements.
				(b) The atoms in the tips of kites are more stable and diffuse slower (note the time stamps) through conventional nearest-neighbor swaps.
				Representative results are for Ta (210), corresponding to \autoref{fig:aperiodic}.
		}}
		\label{fig:wave}
	\end{figure*}

	The generated BCC split kite phases appear to be quasi-aperiodic. 
	Using GRIP, we explore larger area reconstructions of these GBs by tiling the smallest unit cell up to 25 times in the direction along the [001] tilt axis. 
	The nearly degenerate lowest-energy structures for different reconstruction sizes of 1~$\times$~6, 1~$\times$~9 and 1~$\times$~25 are shown in \autoref{fig:aperiodic}a--c. 
	While the structures viewed down the tilt axis appear nearly indistinguishable, the different reconstructions have distinct atomic arrangements along the tilt axis. 
	When viewed from the side, the atoms belonging to the tips of the kites shown in yellow form a periodic wave with a period commensurate with the simulation cell size. 
	\autoref{fig:aperiodic}d shows the GB energy of the different split kite structures as a function of repetition along the tilt axis. 
	The energy converges only for very large reconstructions of 10 and larger, although a structure with a similar low energy can be found for smaller reconstructions (e.g., 1~$\times$~6 for Ta (210)), which we select for subsequent DFT calculations. 
	Even for very large sizes, the energy plot shows oscillations in $E_{\mathrm{gb}}$.
	A similar convergence behavior is observed in the values of $[n]$ for the minimum-energy structure at each reconstruction size in \autoref{fig:aperiodic}d. 
	Additional top views of different cell sizes in the direction orthogonal to the tilt axis are shown in Supp. \autoref{fig:si_aperiodic}.~\cite{supp}

	Split kite phases of $\Sigma$5(210) were identified by empirical potentials in all BCC metals we studied. 
	However, for Mo, W, and Fe, their energies predicted by these models are 5--\SI{8}{\percent} higher than those of the kite ground-state structure. 
	Representative results are shown in \autoref{fig:grip_Mo} for the Mo (210) tilt GB, and similar results for W and Fe are in Supp. \autoref{fig:si_W} and Supp. \autoref{fig:si_Fe}, respectively. 
	Split kites are a metastable phase of this boundary; they are local minima with respect to $[n]$ and remain stable at finite temperatures in MD simulations with periodic boundary conditions (PBCs).

	To validate the stability of the split kite structures predicted at \SI{0}{\kelvin} by GRIP and demonstrate possible GB phase transitions, we perform finite-temperature MD simulations at homologous temperatures ranging from $T_h = 0.2$ to 0.7. 
	To model GB phase transitions from kites to split kites, the GBs were terminated at open surfaces following the methodology proposed in Ref.~\cite{frolov_2013_gbphase}. 
	Open surfaces act as sources and sinks of atoms and effectively introduce a grand-canonical environment in the GB core. 
	First-order phase transitions are observed with open-surface boundary conditions as illustrated in \autoref{fig:grip_NbTa}g and Supp. Movie 1.

	\change{Moreover, the simulations with PBCs reveal that the split kite phases of all metals are dynamically stable at high temperatures. 
		These simulations also reveal that, despite the close energies of the two GB phases, the difference in their atomic structure has a significant impact on diffusivity and the diffusion mechanism.
		\autoref{fig:wave} and Supp. Movie 2 show that the atoms forming the wave-like pattern in the split kites exhibit collective motion, a key component of short-circuit diffusion in GBs~\cite{chesser_2024}.
		The wave is performing a random walk, moving the entire column of atoms and enabling large displacements of atoms along the tilt axis.
		These large displacements seem to be independent of the displacements in adjacent tip columns, as seen in Supp. Movie 3.
		In regular kites, on the other hand, the diffusion mechanism is represented by uncorrelated nearest-neighbor swaps and much slower diffusion (Supp. Movie 4). 
	}

	\begin{table*}
		\renewcommand{\arraystretch}{1.3}
		\caption{\textbf{DFT validation of grain boundary energy.}
			Empirical potential (EP) calculations are performed in GRIP and density functional theory (DFT) calculations are performed in VASP.
			The phase with the lower value of $E_{\mathrm{gb}}$ for each boundary when evaluated with DFT is in bold.
			Units are in \si{\joule\per\meter\squared}.}
		\begin{ruledtabular}
			\begin{tabular}{l|cc|cc|cc|cc|cc|cc|cc}
				\ & \multicolumn{2}{c}{Nb (310)
				} & \multicolumn{2}{c}{Nb (210)
				} & \multicolumn{2}{c}{Ta (210)
				} & \multicolumn{2}{c}{Mo (210)
				} & \multicolumn{2}{c}{W (210)
				} & \multicolumn{2}{c}{Fe (310)
				} & \multicolumn{2}{c}{Fe (210)
				} \\
				\ & kite & split kite & kite & split kite & kite & split kite & kite & split kite & kite & split kite & kite & split kite & kite & split kite \\ \hline
				EP & 0.888 & 0.870 & 0.893 & 0.881 & 1.357 & 1.331 & 1.998 & 2.100 & 2.427 & 2.593 & 1.393 & 1.532 & 1.414 & 1.473 \\
				DFT & \textbf{1.208} & 1.354 & \textbf{1.248} & 1.271 & 1.420 & \textbf{1.390} & 2.001 & \textbf{1.930} & 2.602 & \textbf{2.534} & \textbf{1.540} & 1.632 & \textbf{1.590} & 1.606 \\ 
			\end{tabular}
		\end{ruledtabular}
		\label{tab:dft}
	\end{table*}

	The use of empirical potentials (see Supp. \autoref{tab:si_iap}) enables more thorough structure exploration using GRIP, including the searches of large-area reconstructions that are currently beyond the reach of DFT calculations. 
	Previous studies have demonstrated the accuracy of empirical potentials in predicting new low-energy structures which were subsequently confirmed by DFT and observed in experiments~\cite{frolov_2018, seki_2023}.
	However, it is also well known that predicted GB structures and energies are sensitive to the choice of empirical potential~\cite{mahmood_2024}. 
	We compare the different predictions of empirical potentials in Supp. \autoref{fig:si_Nb_210} and \autoref{fig:si_eam}. 
	With this in mind, we calculate the GB energies of each phase obtained from GRIP using DFT by relaxing a smaller block of atoms extracted from the GB region.
	Additional details on the methodology are provided in the Supplemental Material~\cite{supp}.

	The DFT results summarized in \autoref{tab:dft} confirm split kite phases of $\Sigma$5(210) as the ground state in Ta, Mo, and W, and they have nearly indistinguishable energies from the kite phase for Nb and Fe. 
	The GB energies calculated for the (210) regular kites in this study are consistent with previous reports (Supp. \autoref{tab:si_ref}). 
	None of the regular kite or split kite structures change significantly during the structural relaxation with DFT, and the final structures are presented in Supp. \autoref{fig:si_dft}.

	For split kite phases of the $\Sigma$5(310) boundary, the DFT calculations were only attempted for Nb and Fe, which are the only metals where GRIP searches predicted kite and split kites to have close energies. 
	DFT-predicted split kite GB energies are found to be 6--\SI{12}{\percent} higher than those of kites. 
	We did not explore other metals because the cells would contain over 700 atoms and the calculations would be too computationally demanding.

	
	
	\emph{Discussion}: 
	In this work, we revisit structures of $\Sigma5$[001] symmetric tilt GBs in refractory BCC metals and study their phase behavior using the recently developed GRIP tool~\cite{chen_2024}. 
	Contrary to established and well-known structures, we find new ground states and demonstrate multiple GB phases by optimizing the number of atoms in the boundary core and exploring large-area reconstructions.
	The new BCC split kite phases in [001] tilt GBs can be realized in unit cells of different sizes and periodicities (\autoref{fig:aperiodic}), similar to previously reported split kite structures in [001] tilt GBs in FCC metals~\cite{frolov_2013_gbphase} and other [110] tilt GBs in BCC metals~\cite{frolov_2018, seki_2023}. 
	In this context, our findings provide further evidence that complex aperiodic phases of GBs could be a general phenomenon.

	The BCC metals investigated here have different relative stability between the two GB phases and different ground-state structures. 
	For the reconstructions accessible to DFT, we find that regular kites and split kites have close GB energies, and it is possible that larger area reconstructions could further lower the DFT energies. 
	The existence of metastable split kite phases suggests that under some conditions they may be stabilized by temperature, pressure, or chemical segregation to become ground states~\cite{freitas_2018}.

	\begin{figure}[!ht]     
		\includegraphics[width=\linewidth]{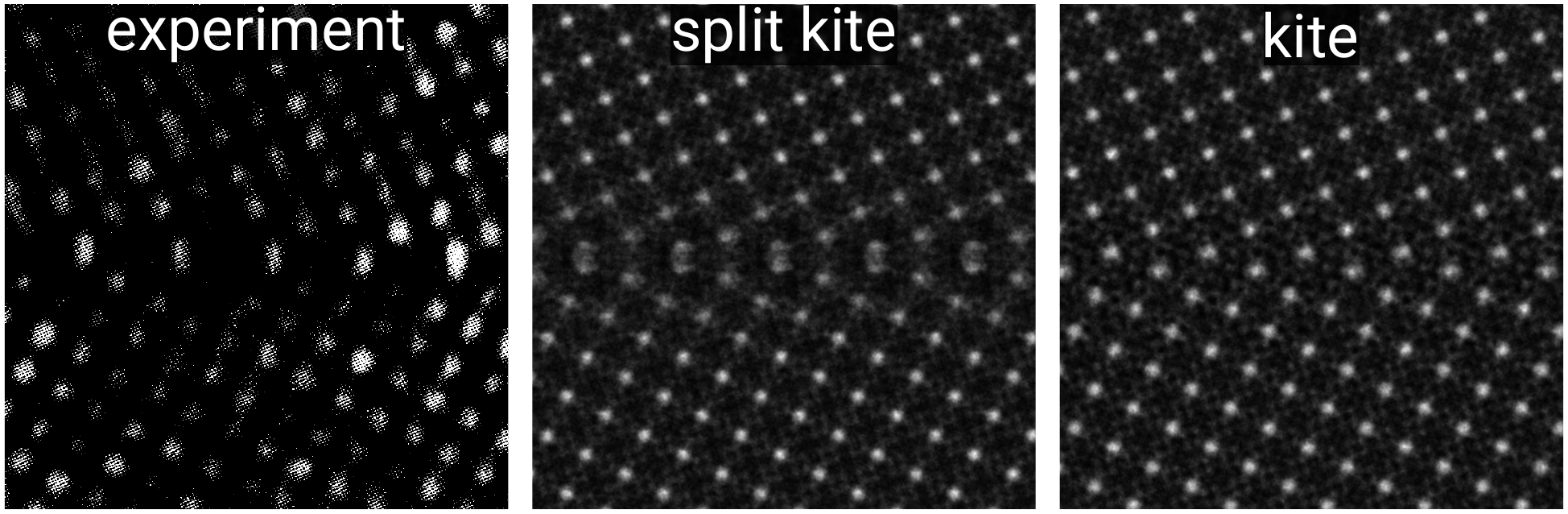}
		\caption{\textbf{HRTEM images of Nb (310).}
			The experimentally observed broadening of the tips of the GB structure is reproduced in the simulated image of the split kite structure, but appears to be absent from the simulated image of the regular kite.
			The experimental image is from Ref.~\cite{campbell_1993} (\textcopyright\ American Physical Society).
		}
		\label{fig:hrtem}
	\end{figure}

	Structures of $\Sigma5$ GBs in several refractory metals were previously systematically investigated by high-resolution TEM (HRTEM) techniques~\cite{campbell_1991, campbell_1993, campbell_1999, campbell_2000, bacia_1997, morita_1997}.
	Unfortunately, in these earlier studies the GB structure was not observed directly. 
	Instead, it was inferred by matching the simulated intensity patterns derived from structures predicted by atomistic modeling with the observed phase contrast in the HRTEM images. 
	To compare our predicted structures with these previous experimental results, we generate the simulated images for split kite and regular kite structures using the multislice method implemented in \emph{ab}TEM~\cite{madsen_2021}.
	The simulated images of the two different structures can be clearly distinguished in \autoref{fig:hrtem} \change{(see Supp. \autoref{fig:si_stem} for HAADF-STEM simulations)}. 
	The wave pattern at the tip of the split kite structure introduces a characteristic broadening in the intensity that is notably absent in the regular kite.
	The same broadening can be seen in the experimental image, suggesting that our predicted split kite structure for Nb (310) may be a better match with previous HRTEM results and may have been experimentally observed.

	Previous experimental studies have also reported rigid body translations of (002) across the GB plane along the [001] tilt axis~\cite{campbell_2002}.
	We report similar displacements for kites in Nb, Ta, and Mo in Supp. \autoref{tab:si_shift}, consistent with some~\cite{ochs_2000a, scheiber_2016} but not all~\cite{zheng_2020} \emph{ab initio} studies.
	Notably, the lack of translations in our simulated Nb (310) split kite structure is also in agreement with experimental reports~\cite{campbell_1993}.
	Given the sensitivity of the simulated structures and images to the simulation parameters, our results motivate further experimental studies using modern aberration-corrected electron microscopes that can directly resolve atomic positions.
	Exact mechanisms leading to the formation of complex GB structures are currently unknown and warrant further investigation.


	\begin{acknowledgments}
		This work was performed under the auspices of the U.S. Department of Energy (DOE) by the Lawrence Livermore National Laboratory (LLNL) under Contract No. DE-AC52-07NA27344. 
		The authors thank Pedro Borges, Colin Ophus, and Omar Hussein for helpful discussions.
		Computing support for this work comes from the LLNL Institutional Computing Grand Challenge program.
		An award of computer time was also provided by the INCITE program. 
		This research used resources of both the Argonne and Oak Ridge Leadership Computing Facilities, which are DOE Office of Science User Facilities supported under contracts DE-AC02-06CH11357 and DE-AC05-00OR22725, respectively.
		T.F. acknowledges support from the U.S. DOE, Office of Science under an Office of Fusion Energy Sciences Early Career Award.
		All figures are produced using matplotlib~\cite{hunter_2007} and OVITO~\cite{stukowski_2009}. 
	\end{acknowledgments}
	
	\emph{Author contributions}:
	T.F. conceptualized and supervised the project.
	E.C. performed the simulations.
	Both authors analyzed the results and wrote the manuscript.
	The authors declare no competing financial or non-financial interests.
	
	\emph{Data availability}:
	The {GRand} canonical Interface Predictor (GRIP) tool that generated the data for this work can be found at \url{https://github.com/enze-chen/grip}.
	The data that support the findings of this article are openly available~\cite{zenodo}.
	
	

\bibliography{main}

\newpage
\onecolumngrid

\include{supp0.tex}

\end{document}

%% file: supp0.tex
\begin{center}
    {\large  
    \textbf{Supplementary Information for \\ \vspace{0.05in}
    Quasi-aperiodic grain boundary phases of $\Sigma5$ tilt grain boundaries in refractory metals}
    \vspace{0.1in}}
    
    {Enze Chen$^{1, *}$ and Timofey Frolov$^{2, \dagger}$}
\end{center}

{\noindent \fontsize{10}{12}\selectfont 
$^1$ Department of Materials Science and Engineering, Stanford University, Stanford, CA 94305, USA \\
$^2$ Materials Science Division, Lawrence Livermore National Laboratory, Livermore, CA 94550, USA \\
Email: 
$^{*}$enze@stanford.edu; 
$^{\dagger}$frolov2@llnl.gov
}

\setcounter{figure}{0}
\renewcommand{\thefigure}{S\arabic{figure}}
\renewcommand{\figurename}{SUPPLEMENTARY FIG.}
\setcounter{table}{0}
\renewcommand{\thetable}{S\arabic{table}}
\renewcommand{\tablename}{SUPPLEMENTARY TABLE}
\setcounter{equation}{0}
\renewcommand{\theequation}{S\arabic{equation}}
\vspace{16pt}

\begin{table}[!ht]
    \renewcommand{\arraystretch}{1.3}
    \caption{\textbf{Empirical interatomic potentials used in this work.}
    EAM = embedded-atom method~\cite{daw_1984}, MEAM = modified embedded-atom method~\cite{baskes_1992}, ADP = angular-dependent potential~\cite{mishin_2005}.
    }
    \begin{ruledtabular}
    \begin{tabular}{c|ccccc}
        \ & Nb & Ta & Mo & W & Fe \\ \hline
        EAM & Fellinger et al.~\cite{fellinger_2010} & Li et al.~\cite{li_2003} & Zhou et al.~\cite{zhou_2004} & Marinica et al.~\cite{marinica_2013} & Proville et al.~\cite{proville_2012} \\
        MEAM & Yang and Qi~\cite{yang_2019} & & Park et al.~\cite{park_2012} & Hiremath et al.~\cite{hiremath_2022} & \\
        ADP & & Purja Pun et al.~\cite{purjapun_2015} & & & Starikov et al.~\cite{starikov_2021} \\
    \end{tabular}
    \end{ruledtabular}
    \label{tab:si_iap}
\end{table}

\vfill

\begin{table}[!ht]
    \renewcommand{\arraystretch}{1.3}
    \caption{\textbf{Comparison of calculated GB energies with literature values.}
    Only the energies of kite structures are compared with those from the literature, though our DFT-calculated energies for split kites (SK) are also listed.
    EP = empirical potential and DFT = density functional theory.
    The values are consistent for all boundaries, especially for DFT calculations.
    Units are in \si{\joule\per\meter\squared}.}
    \begin{ruledtabular}
    \begin{tabular}{c|cc|cc|cc|cc|cc|cc|cc}
        \ & \multicolumn{2}{c}{Nb (310)
        } & \multicolumn{2}{c}{Nb (210)
        } & \multicolumn{2}{c}{Ta (210)
        } & \multicolumn{2}{c}{Mo (210)
        } & \multicolumn{2}{c}{W (210)
        } & \multicolumn{2}{c}{Fe (310)
        } & \multicolumn{2}{c}{Fe (210)
        } \\
        \ & kite & ref. & kite & ref. & kite & ref. & kite & ref. & kite & ref. & kite & ref. & kite & ref. \\ \hline
        EP & 0.888 & 1.117~\cite{singh_2020} & 0.893 & 1.257~\cite{singh_2020} & 1.357 & 1.351~\cite{hahn_2016} & 1.998 & 2.032~\cite{waters_2023} & 2.427 & 2.500~\cite{frolov_2018_W} & 1.393 & 1.442~\cite{starikov_2021} & 1.414 & 1.451~\cite{starikov_2021} \\
        DFT & 1.208 & 1.20~\cite{zheng_2020} & 1.248 & 1.24~\cite{zheng_2020} & 1.420 & 1.41~\cite{zheng_2020} & 2.001 & 2.03~\cite{zheng_2020} & 2.602 & 2.65~\cite{zheng_2020} & 1.540 & 1.565~\cite{wang_2018} & 1.590 & 1.64~\cite{wang_2018} \\ \hline 
        SK & 1.354 & --- & 1.271 & --- & 1.390 & --- & 1.930 & --- & 2.534 & --- & 1.632 & --- & 1.606 & --- \\
    \end{tabular}
    \end{ruledtabular}
    \label{tab:si_ref}
\end{table}

\vfill

\begin{table}[!ht]
    \renewcommand{\arraystretch}{1.3}
    \caption{\textbf{Energy of shifted kite structures.}
    Shown below are only the GB structures where GRIP predicts a rigid body translation of (002) across the GB along the [001] tilt axis.
    Results from relaxations with empirical potentials (EP) are compared with those from density functional theory (DFT) for shifted and non-shifted GB structures.
    Units are in \si{\joule\per\meter\squared}.}
    \begin{ruledtabular}
    \begin{tabular}{llllll}
        \ & Nb (310) & Nb (210) & Ta (310) & Mo (310) & Mo (210) \\ \hline 
        EP relax (shift) & 0.888 & 0.893 & 1.315 & 1.743 & 1.998 \\
        DFT relax (shift) & 1.208 & 1.248 & 1.457 & 1.705 & 2.001 \\
        DFT relax (no shift) & 1.188 & 1.262 & 1.453 & 1.810 & 2.066 \\
        Crystallium (no shift)~\cite{zheng_2020} & 1.20 & 1.24 & 1.44 & 1.73 & 2.03
    \end{tabular}
    \end{ruledtabular}
    \label{tab:si_shift}
\end{table}

\newpage 
\begin{figure}[!h]
    \includegraphics[width=0.95\linewidth]{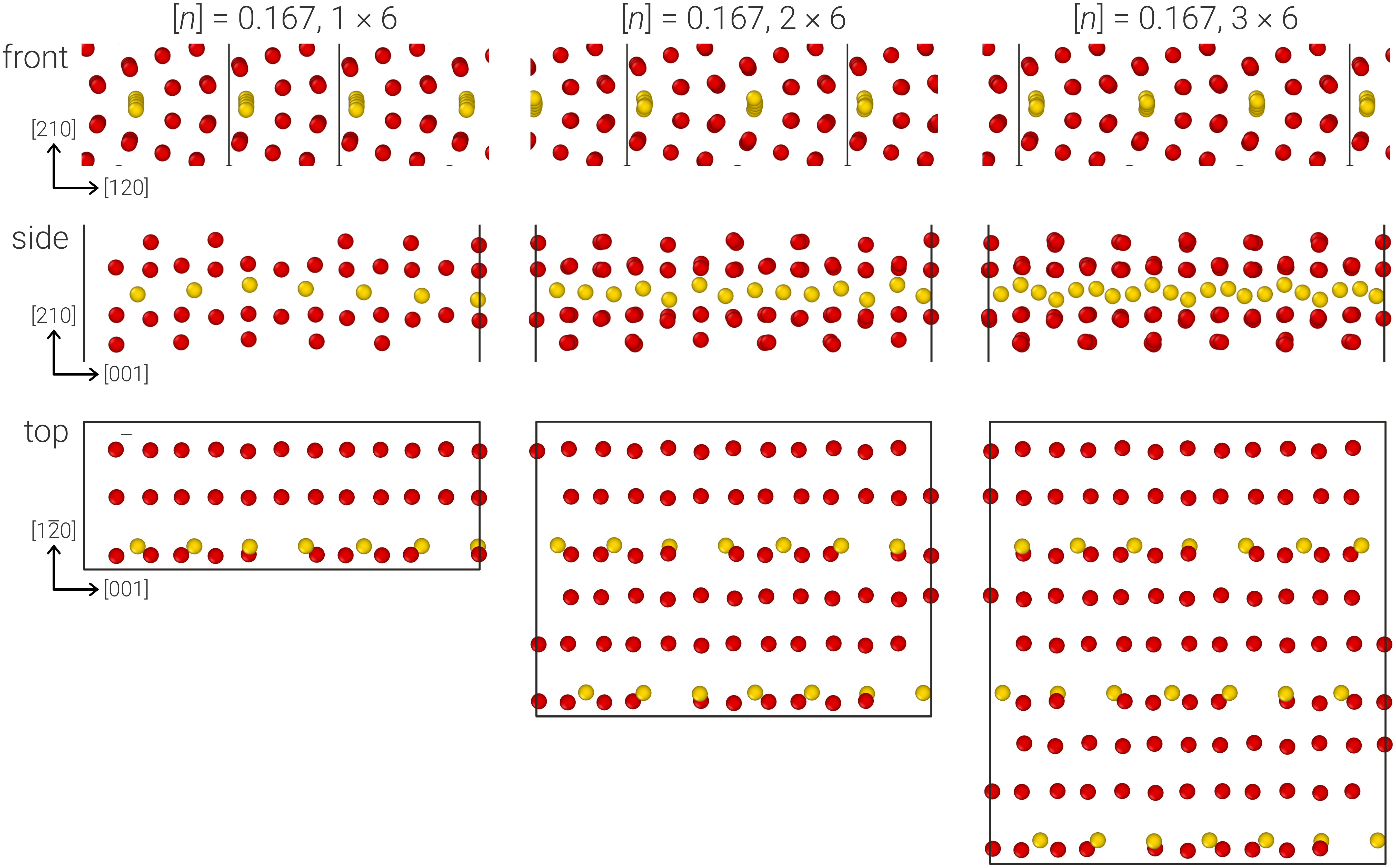}
    \caption{\textbf{Quasi-aperiodicity of Ta (210) orthogonal to the [001] tilt axis.}
    For larger reconstructions orthogonal to the tilt axis (along $[1\bar{2}0]$), the split kite phase also forms distinct wave-like patterns that are shifted varying amounts along $[001]$.
    All minimum-energy structures are visually similar when viewed from the front and have the same GB atomic density of $[n]=1/6$.
    }
    \label{fig:si_aperiodic}
\end{figure}

\newpage 
\begin{figure}[!h]
    \includegraphics[width=\linewidth]{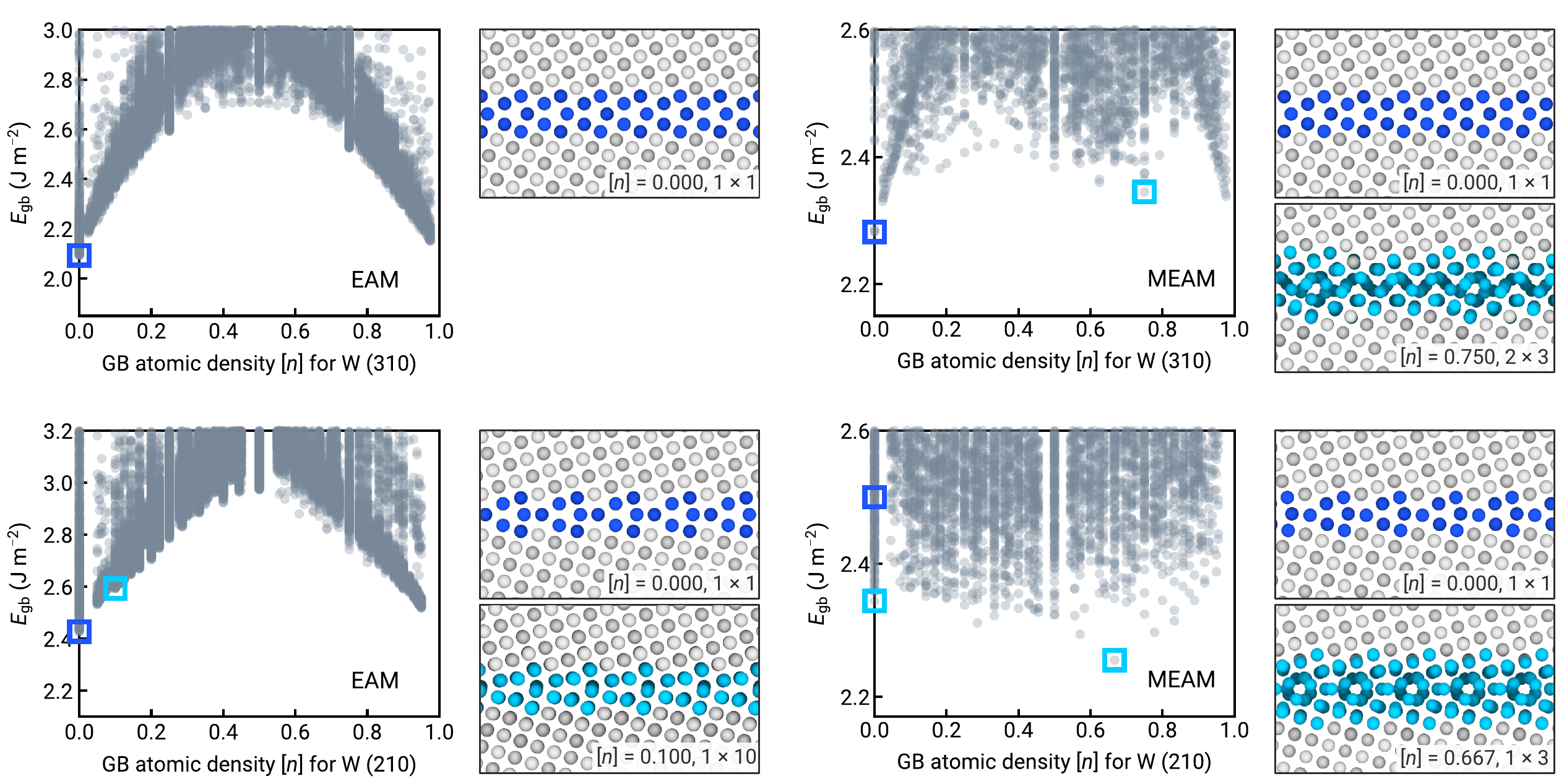}
    \caption{\textbf{GRIP results for $\Sigma5$[001] GBs in W.}
    Similar to Mo (\autoref{fig:grip_Mo} in the main manuscript), the split kite structure is not the ground state in W when evaluated with an empirical potential~\cite{marinica_2013}, and only appears for (210) at $[n]=0.1$ (\textcolor{DeepSkyBlue}{light blue}, lower left).
    GB atoms are colored to highlight their distinct coordination.
    }
    \label{fig:si_W}
\end{figure}

\newpage 
\begin{figure}[!h]
    \includegraphics[width=0.9\linewidth]{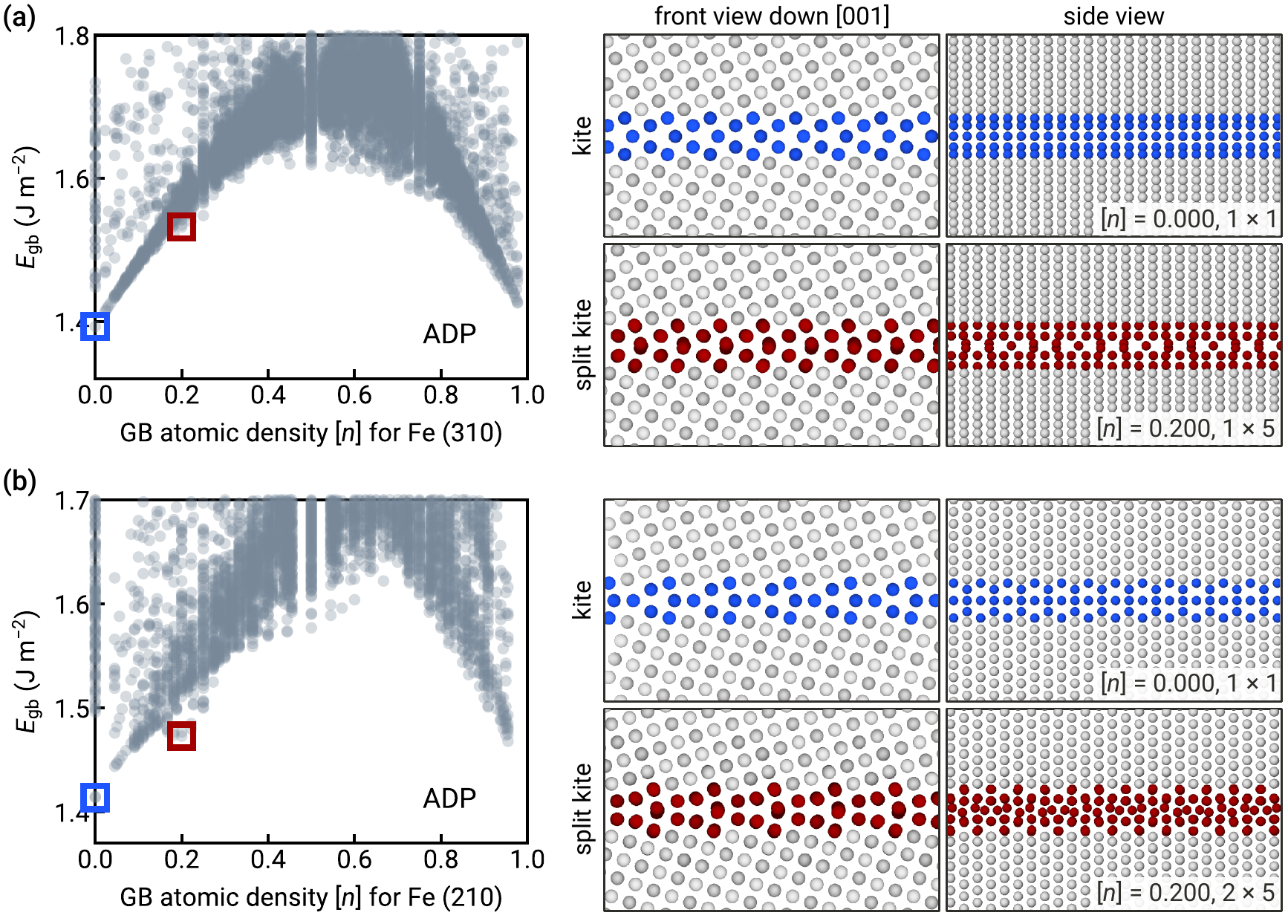}
    \caption{\textbf{Complex split kite phases in $\Sigma5$[001] GBs in Fe.}
    Similar to Mo (\autoref{fig:grip_Mo} in the main manuscript), the split kite structure is metastable in (a) Fe (310) and (b) Fe (210) when evaluated with an empirical potential~\cite{purjapun_2015}.
    GB atoms are colored to highlight their distinct coordination in the kite (\textcolor{blue}{blue}) and split kite (\textcolor{Maroon}{maroon}) phases.
    }
    \label{fig:si_Fe}
\end{figure}


\newpage 
\begin{figure}[!h]
    \includegraphics[width=0.9\linewidth]{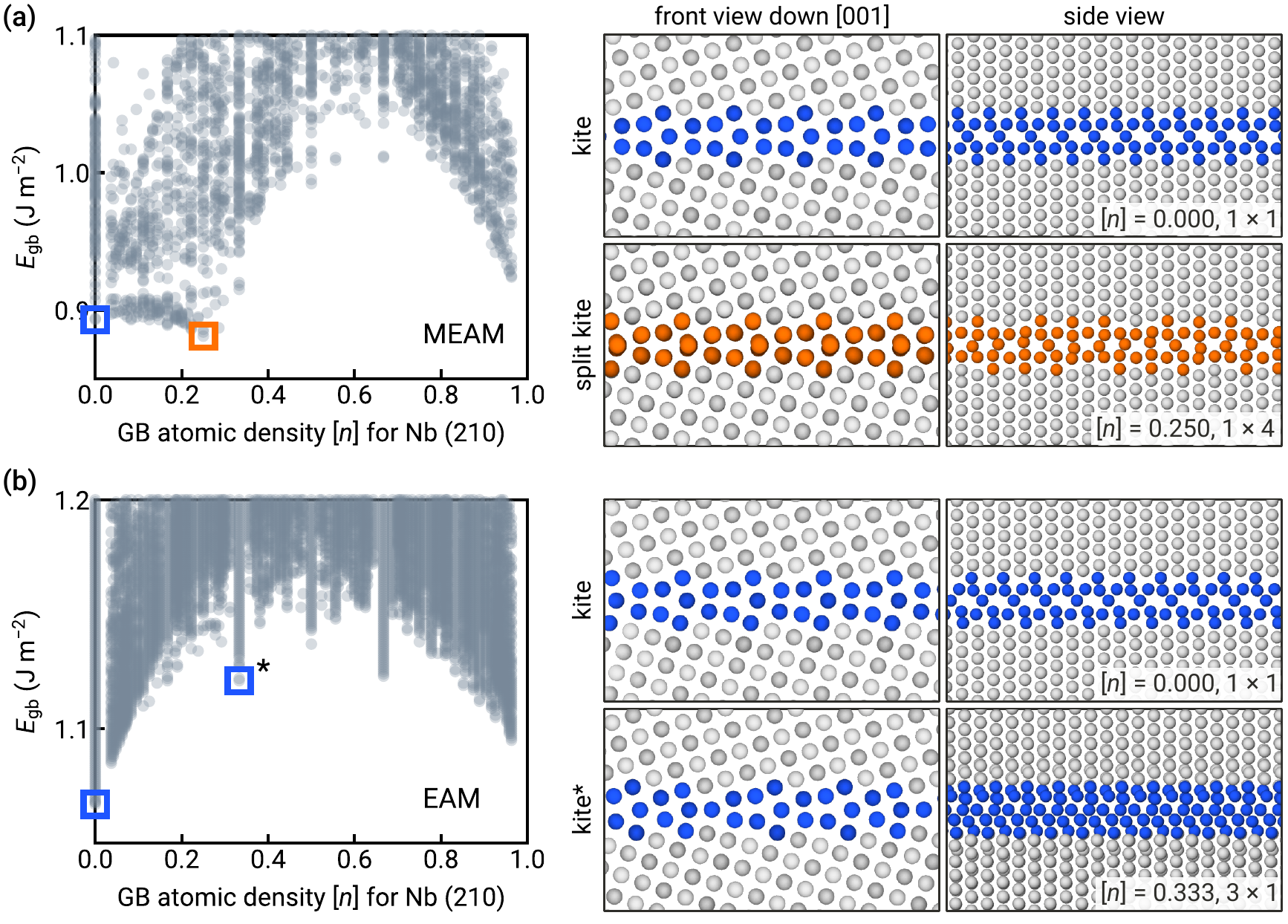}
    \caption{\textbf{GRIP results for $\Sigma5(210)[001]$ in Nb.}
    (a) When modeled with a MEAM potential~\cite{yang_2019}, the split kite phase is the ground state in Nb (210). 
    (b) When modeled with an EAM potential~\cite{fellinger_2010}, the split kite phase is not a minimum-energy structure.
    GB atoms are colored to highlight their distinct coordination in the kite (\textcolor{blue}{blue}) and split kite (\textcolor{DarkOrange}{orange}) phases.
    }
    \label{fig:si_Nb_210}
\end{figure}

\newpage 
\begin{figure}[!h]
    \includegraphics[width=0.76\linewidth]{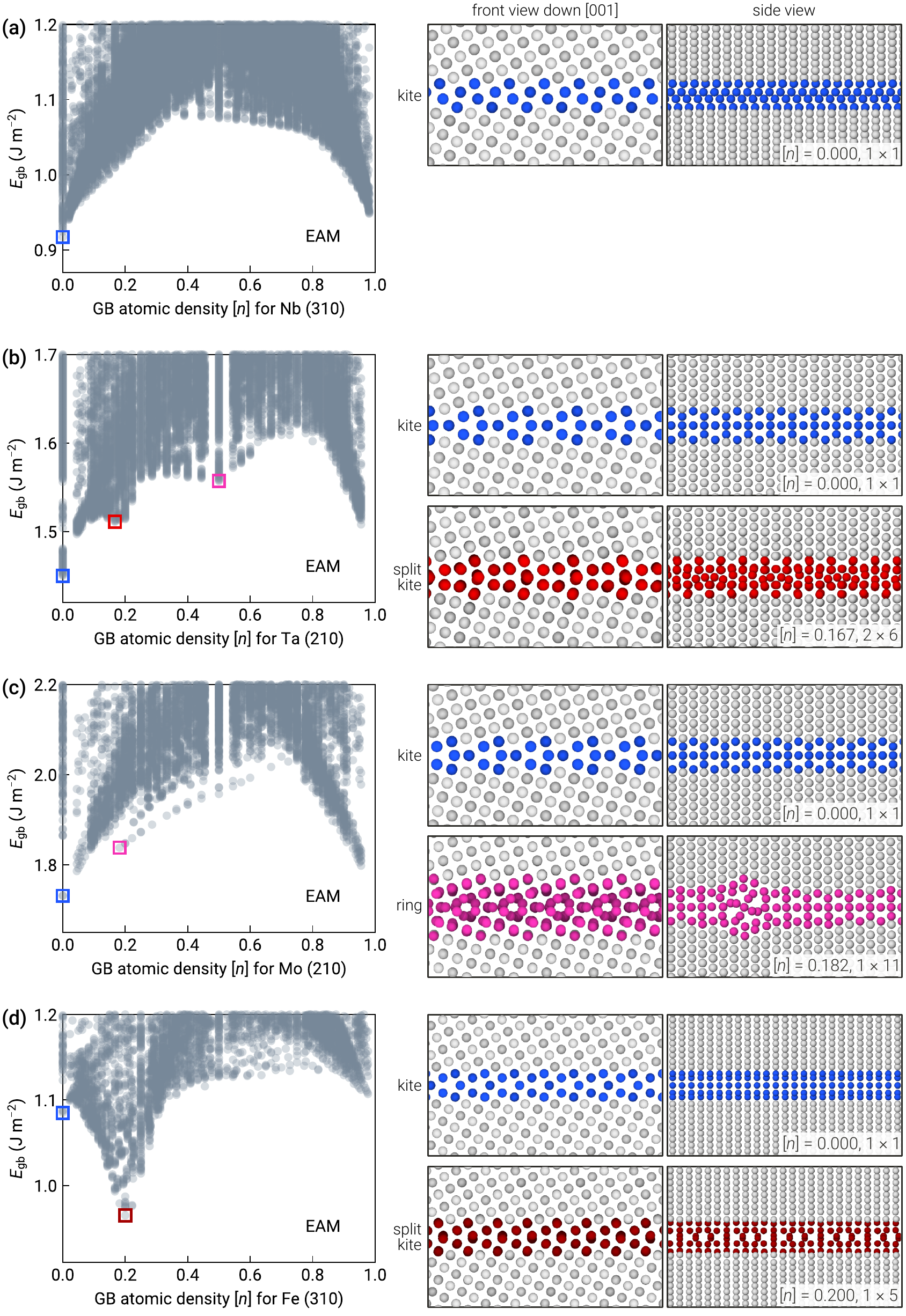}
    \caption{\textbf{GRIP results using embedded-atom method (EAM) potentials for $\Sigma5[001]$ GBs.}
    Performing a structure search using GRIP with EAM potentials~\cite{fellinger_2010, li_2003, zhou_2004, proville_2012} produces dramatically different results. 
    (a) The split kite phase is no longer a minimum-energy structure in Nb (310). 
    (b) The split kite phase is a metastable phase in Ta (210). 
    (c) The split kite phase is no longer a minimum-energy structure in Mo (210).
    (d) The split kite phase is the ground state in Fe (310).
    GB atoms are colored to highlight their distinct coordination.
    }
    \label{fig:si_eam}
\end{figure}

\newpage 
\begin{figure}[!h]
    \includegraphics[width=0.65\linewidth]{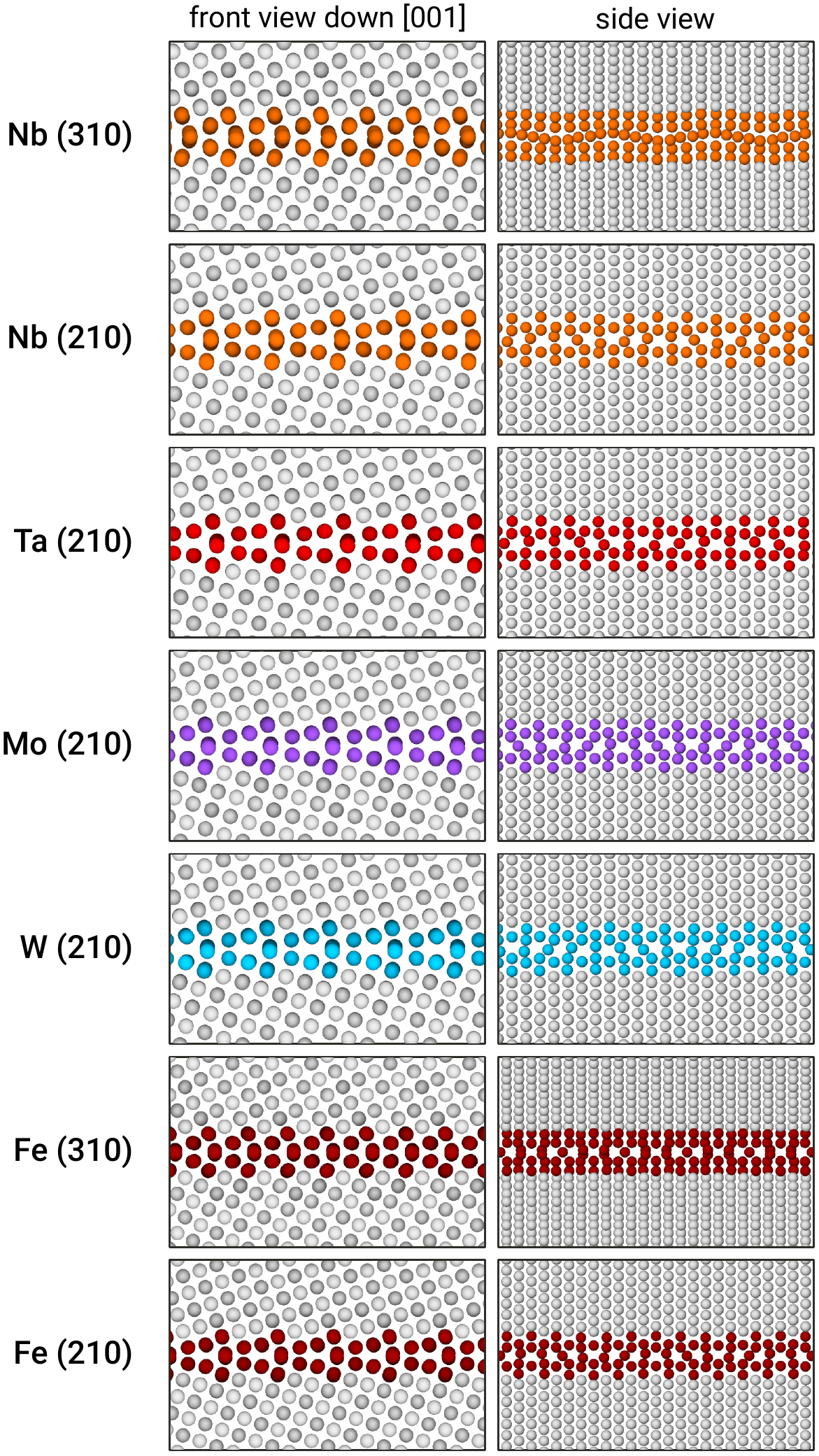}
    \caption{\textbf{Split kite structures relaxed using density functional theory (DFT) calculations.}
    DFT calculations confirm that the split kite structures are metastable and have GB energies that are competitive with those of regular kites (\autoref{tab:dft} in the main manuscript) in all of the boundaries shown here.
    GB atoms are colored to highlight their distinct coordination with colors consistent to those of split kites from GRIP.
    }
    \label{fig:si_dft}
\end{figure}

\newpage 
\begin{figure}[!h]
	\includegraphics[width=\linewidth]{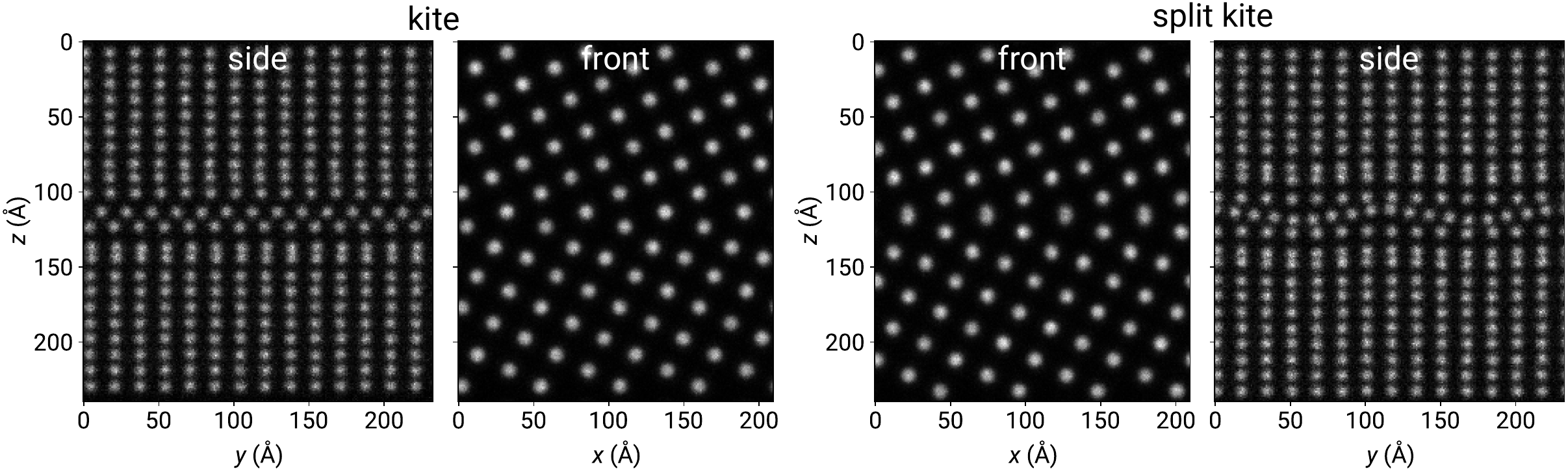}
	\caption{\change{\textbf{HAADF-STEM image simulations for Nb (310).}
			In addition to HRTEM image simulations presented in \autoref{fig:hrtem} in the main manuscript, we performed additional HAADF-STEM image simulations using \emph{ab}TEM~\cite{madsen_2021}.
			The slight blurring of the kite tips (front view) and the wave pattern (side view) are clearly distinguishable in the split kite phase.
			We adopt the parameters reported in Ref.~\cite{zhou_2023}, namely $E = \SI{300}{\kilo\volt}$ and $\alpha = \SI{23.6}{\milli\radian}$.
			We note these settings should be regarded as only representative as no STEM studies for Nb (310) have been reported in the literature, and we caution that these subtle features may easily be misinterpreted.}
	}
	\label{fig:si_stem}
\end{figure}

\vspace{8ex}
\section*{SUPPLEMENTARY MOVIE 1}

Video of a first-order phase transformation in the Ta (210) GB at \SI{1500}{\kelvin}, where the regular kites transform into split kites at approximately 00:21 in the movie.
The MD simulation is performed with open surfaces on both sides and each snapshot is relaxed for clarity of visualization.
The movies can be found with the published manuscript or in Google Drive (\href{https://drive.google.com/file/d/1m70srFysVMECkWBvFAIZyLK5yFj1t6CS/view?usp=sharing}{relaxed}, \href{https://drive.google.com/file/d/1Frtv6yuAC0r46BWw6e350QQE46OyzPH5/view?usp=sharing}{unrelaxed}).

\change{
\section*{SUPPLEMENTARY MOVIE 2}

Video of the side view (looking down $[1 \bar{2} 0]$) of one atomic column of the tips of split kites.
The quasi-aperiodic wave pattern persists at high temperature.
A single atom is colored blue (tracer) to track the dynamics, which feature collective diffusion back and forth along the atomic column.
Each frame is \SI{1}{\nano\second} and the whole video is \SI{40}{\nano\second}.
The movie can be found with the published manuscript or \href{https://drive.google.com/file/d/1Mc9gOEcomiiYN_v55D-hsDlDE8_zRDc1/view?usp=drive_link}{in Google Drive}.

\section*{SUPPLEMENTARY MOVIE 3}

Video of the side view (looking down $[1 \bar{2} 0]$) of two adjacent atomic columns of the tips of split kites.
A single atom is colored in each atomic column (blue in the yellow wave and magenta in the green wave) to track the independent wave oscillations in separate GB cores.
Each frame is \SI{1}{\nano\second} and the whole video is \SI{40}{\nano\second}.
The movie can be found with the published manuscript or \href{https://drive.google.com/file/d/1qx-2PkoHBh-0GlWqvUwxAnyLwmfPBYO5/view?usp=drive_link}{in Google Drive}.

\section*{SUPPLEMENTARY MOVIE 4}

Video of the side view (looking down $[1 \bar{2} 0]$) of one atomic column of the tips of kites.
A single atom is colored blue to track the dynamics, which are slow and feature uncorrelated nearest-neighbor hops.
Each frame is \SI{1}{\nano\second} and the whole video is \SI{40}{\nano\second}.
The movie can be found with the published manuscript or \href{https://drive.google.com/file/d/1W4YiHGyix9vnmJX3BpOvzl0ZVzzhxm_O/view?usp=drive_link}{in Google Drive}.
}

\section*{SUPPLEMENTARY METHODS}

\subsection*{Grand canonical Interface Predictor (GRIP)}

We perform atomic-level optimization of GB structures using the open-source, Python-based {GRand} canonical Interface Predictor (GRIP) code~\cite{chen_2024}, which rigorously explores structural degrees of freedom (DOF) through dynamic sampling.
Bicrystal slabs are automatically generated in the code using the Atomic Simulation Environment (ASE)~\cite{larsen_2017} library, and the orthogonal simulation is oriented such that the \mbox{$y$-axis} is the tilt axis direction, the \mbox{$x$-axis} is the orthogonal in-plane direction, and the \mbox{$z$-axis} is the out-of-plane normal direction.
We ensure periodicity in the GB plane (\mbox{$xy$-plane}) and at least \SI{3.5}{\nano\meter} in the \mbox{$z$-direction} for each slab to minimize cell size effects.
There is only one GB in the middle of the simulation cell due to aperiodic boundary conditions in the \mbox{$z$-direction.}

For an individual GB, each iteration of the algorithm has three stages.
During the first stage, the initial configuration is created by uniformly sampling a set of parameters specified through the input file. 
In this study, the algorithm randomly samples a $3\, \times\, 9$ replication of the unit GB cell, randomly translates the upper slab in the \mbox{$xy$-plane}, and removes a randomly chosen fraction of atoms from the GB. 
The simulation box size scales with the replications to maintain in-plane periodicity.

During the second stage, it performs dynamic sampling to optimize the GB structure consistent with the imposed DOF. 
In this study, we use standard finite-temperature MD simulations using the Large-scale Atomic/Molecular Massively Parallel Simulator (LAMMPS)~\cite{thompson_2022} in the canonical ($NVT$) ensemble with a Langevin thermostat and a time step of \SI{2}{\femto\second} as our dynamic sampling technique.
The temperature and duration of the MD are also randomly sampled and only the atoms in a region \SI{1}{\nano\meter} above and below the GB plane (referred to as the ``GB region") are allowed to move freely during the dynamic sampling phase. 
For this study, we sample a temperature range between 0.2 and 0.7 of the melting temperature of each metal, and a duration up to \SI{0.4}{\nano\second}.
Finally, the temperature is quickly ramped down to \SI{100}{\kelvin} for \SI{2}{\pico\second}.
Additionally, there is a \SI{5}{\percent} chance the algorithm skips the dynamic sampling for one iteration and jumps to the third stage.

In the third stage, each GB structure is fully relaxed at \SI{0}{\kelvin} using a conjugate gradient minimization scheme, where atoms in the GB and buffer regions (an addition \SI{1}{\nano\meter} on each side of the GB region) can move freely.
The convergence criteria are \SI{e-15} for relative energy ($\mathrm{d}E/E$ in successive iterations) and \SI{e-15}{\electronvolt\per\angstrom} for forces, with a maximum of \SI{e5} evaluations for each criterion.
We found this strict criteria necessary for full convergence of the GB structures, particularly when the two grains are displaced along the tilt axis.
The algorithm repeats these stages on each processor independently until termination, saving each relaxed structure to disk and periodically deleting duplicates.
Duplicates are defined as structures with the same value of $E_{\mathrm{gb}}$ and $[n]$ to three decimal places, and the algorithm will keep the structure with a smaller reconstruction and relative translations.

For each relaxed structure, the GB energy, $E_{\mathrm{gb}}$, is computed according to:

\begin{equation*}
    E_{\mathrm{gb}} = \frac{E^{\mathrm{gb}}_{\mathrm{total}} - N^{\mathrm{gb}}_{\mathrm{total}} E^{\mathrm{bulk}}_{\mathrm{coh}}}{A^{\mathrm{gb}}_{\mathrm{plane}}}
\end{equation*}

\noindent as explained in \autoref{eq:Egb} in the Results section of the main manuscript.
We also track the fraction of atoms in one plane or GB atomic density, $[n]$, according to:

\begin{equation*}
    [n] = \frac{N_{\mathrm{gb}} \mod N_{\mathrm{plane}}}{N_{\mathrm{plane}}} \in [0, 1)
\end{equation*}

\noindent as explained in \autoref{eq:n} in the Results.
Multiple interatomic potential parameterizations are used for each system, as shown in \autoref{tab:si_iap}.

\newpage 
\subsection*{Density functional theory (DFT)}

To validate select GB structures, we perform additional density functional theory (DFT) calculations using the Vienna Ab initio Simulation Package (VASP)~\cite{kresse_1993, kresse_1994, kresse_1996a, kresse_1996b} with projector augmented-wave potentials~\cite{kresse_1999} and the generalized gradient approximation exchange correlation functional of Perdew, Burke, and Ernzerhof~\cite{perdew_1996}.
The semi-core $3p$ states are treated as valence states for Nb, Ta, and Mo (\texttt{*\_pv} potential), while Fe and W use the standard potentials.
Calculations for Fe are spin-polarized with initial magnetic moments of \SI{3.0}{\muB}.
We use Monkhorst-Pack~\cite{monkhorst_1976} $\mathbf{k}$-point grids with a density of 7000 $\mathbf{k}$ points per reciprocal atom (calculated using Pymatgen~\cite{ong_2013}) and apply Methfessel-Paxton smearing~\cite{methfessel_1989} with a width of \SI{0.1}{\electronvolt}.
The plane wave cutoff energy is \SI{500}{\electronvolt} and the convergence criteria are set at \SI{e-5}{\electronvolt} for energy and \SI{0.02}{\electronvolt\per\angstrom} for forces.

We create the input structure by extracting a section near the GB region of the optimized structure from GRIP of approximately \SI{6}{\nano\meter} in thickness (except for Nb (310), which is \SI{5}{\nano\meter} thick) and adding \SI{1}{\nano\meter} of vacuum on top. 
This results in computational cells ranging from 80 atoms (Ta (210) regular kites) up to 662 atoms (Nb (310) split kites).
The axes are rescaled to match equilibrium DFT values and atomic positions are fully relaxed while the cell shape and volume are fixed to maintain stresses in the GB plane.
The energy of the GB is computed as the difference in total energy of a structure with the GB and a second bulk structure in the same orientation with the same number of atoms and vacuum but without a GB (i.e., a bulk slab), divided by the planar area.
This subtraction aims to eliminate the contribution from the two free surfaces in the periodic cell.
When the number of atoms between the two structures (GB and bulk) differ, as for split kites, we adjust the energy of the bulk slab by subtracting a DFT-calculated cohesive energy per atom multiplied by the difference in atoms.
Mathematically, the GB energy from DFT may be expressed as:

\begin{equation*}
    E_{\mathrm{gb}}^{\textsc{dft}} = \frac{E^{\mathrm{gb-cell}}_{\mathrm{total}} - \left( E^{\mathrm{bulk-cell}}_{\mathrm{total}} - \Delta N^{\mathrm{bulk-gb}} E^{\textsc{dft}}_{\mathrm{coh}}\right) }{A^{\mathrm{gb}}_{\mathrm{plane}}}    
\end{equation*}

\vfill 
\subsection*{Molecular dynamics (MD)}

To study GB phase stability and transitions, we perform finite-temperature MD simulations using methods adapted from previous work~\cite{chen_2024}.
Briefly, we replicate the optimized GB structures in the $x$- and $y$-directions until the simulation cell is around \SI{30}{\nano\meter} in the $x$-direction and \SI{10}{\nano\meter} in the $y$-direction along the tilt axis.
We freeze the bottom \SI{1}{\nano\meter} layer of atoms and constrain the top \SI{1}{\nano\meter} layer to be semi-rigid throughout the simulation, which lasts at least \SI{20}{\nano\second}.
We use periodic boundary conditions (PBCs) in the $y$-direction and both PBCs and open surfaces with \SI{1}{\nano\meter} of vacuum in the $x$-direction.
We scan a range of temperatures between 0.2 and 0.7 of the melting temperature of each system.
For clarity of visualization, we relax all structures at \SI{0}{\kelvin} using a conjugate gradient minimization scheme.
\change{In Figure 4, the atomic positions are time-averaged across 1500 time steps (\SI{3}{\pico\second}) instead to preserve some high-temperature behavior.}

\vfill 
\subsection*{Image simulations}

To compare our results to previous experimental studies that used high-resolution transmission electron microscopy (HRTEM), we perform phase-contrast image simulations using the multislice algorithm implemented in the \emph{ab}TEM code~\cite{madsen_2021}.
The optimized grain boundary structures from GRIP for Nb (310) are used as input structures, replicated as necessary to achieve a lateral distance of at least \SI{60}{\nano\meter} and a thickness of at least \SI{70}{\nano\meter} along the tilt axis.
We use 64 frozen phonons to approximate atomic vibrations~\cite{loane_1991} with a standard deviation of atomic displacements of \SI{0.1}{\angstrom} and the infinite potential parametrization from Kirkland~\cite{kirkland_2020}.
The electron beam energy is set at \SI{300}{\kilo\volt} and the contrast transfer function parameters are $C_{1,0} = \SI{11.5}{\nano\meter}$ (defocus of \SI{-11.5}{\nano\meter}), $C_{3,0} = -\SI{45}{\micro\meter}$ (spherical aberration), and $C_{5,0} = \SI{5}{\milli\meter}$.
\change{For high-angle annular dark-field STEM image simulations, the same electron beam energy was used with a semi-converge angle of \SI{23.6}{\milli\radian}, following Ref.~\cite{zhou_2023}.}




%% file: main.bbl
\begin{thebibliography}{66}%
\makeatletter
\providecommand \@ifxundefined [1]{%
 \@ifx{#1\undefined}
}%
\providecommand \@ifnum [1]{%
 \ifnum #1\expandafter \@firstoftwo
 \else \expandafter \@secondoftwo
 \fi
}%
\providecommand \@ifx [1]{%
 \ifx #1\expandafter \@firstoftwo
 \else \expandafter \@secondoftwo
 \fi
}%
\providecommand \natexlab [1]{#1}%
\providecommand \enquote  [1]{``#1''}%
\providecommand \bibnamefont  [1]{#1}%
\providecommand \bibfnamefont [1]{#1}%
\providecommand \citenamefont [1]{#1}%
\providecommand \href@noop [0]{\@secondoftwo}%
\providecommand \href [0]{\begingroup \@sanitize@url \@href}%
\providecommand \@href[1]{\@@startlink{#1}\@@href}%
\providecommand \@@href[1]{\endgroup#1\@@endlink}%
\providecommand \@sanitize@url [0]{\catcode `\\12\catcode `\$12\catcode
  `\&12\catcode `\#12\catcode `\^12\catcode `\_12\catcode `\%12\relax}%
\providecommand \@@startlink[1]{}%
\providecommand \@@endlink[0]{}%
\providecommand \url  [0]{\begingroup\@sanitize@url \@url }%
\providecommand \@url [1]{\endgroup\@href {#1}{\urlprefix }}%
\providecommand \urlprefix  [0]{URL }%
\providecommand \Eprint [0]{\href }%
\providecommand \doibase [0]{https://doi.org/}%
\providecommand \selectlanguage [0]{\@gobble}%
\providecommand \bibinfo  [0]{\@secondoftwo}%
\providecommand \bibfield  [0]{\@secondoftwo}%
\providecommand \translation [1]{[#1]}%
\providecommand \BibitemOpen [0]{}%
\providecommand \bibitemStop [0]{}%
\providecommand \bibitemNoStop [0]{.\EOS\space}%
\providecommand \EOS [0]{\spacefactor3000\relax}%
\providecommand \BibitemShut  [1]{\csname bibitem#1\endcsname}%
\let\auto@bib@innerbib\@empty
\bibitem [{\citenamefont {Sutton}\ and\ \citenamefont
  {Balluffi}(1995)}]{sutton_1995}%
  \BibitemOpen
  \bibfield  {author} {\bibinfo {author} {\bibfnamefont {A.~P.}\ \bibnamefont
  {Sutton}}\ and\ \bibinfo {author} {\bibfnamefont {R.~W.}\ \bibnamefont
  {Balluffi}},\ }\href
  {https://global.oup.com/academic/product/interfaces-in-crystalline-materials-9780199211067}
  {\emph {\bibinfo {title} {Interfaces in Crystalline Materials}}},\ \bibinfo
  {edition} {1st}\ ed.\ (\bibinfo  {publisher} {Oxford University Press},\
  \bibinfo {address} {New York},\ \bibinfo {year} {1995})\BibitemShut {NoStop}%
\bibitem [{\citenamefont {Ranganathan}(1966)}]{ranganathan_1966}%
  \BibitemOpen
  \bibfield  {author} {\bibinfo {author} {\bibfnamefont {S.}~\bibnamefont
  {Ranganathan}},\ }\bibfield  {title} {\bibinfo {title} {On the geometry of
  coincidence-site lattices},\ }\href
  {https://doi.org/10.1107/S0365110X66002615} {\bibfield  {journal} {\bibinfo
  {journal} {Acta Crystallographica}\ }\textbf {\bibinfo {volume} {21}},\
  \bibinfo {pages} {197} (\bibinfo {year} {1966})}\BibitemShut {NoStop}%
\bibitem [{\citenamefont {Ding}\ \emph {et~al.}(2025)\citenamefont {Ding},
  \citenamefont {Akbari}, \citenamefont {Chen}, \citenamefont {R{\"o}sner},
  \citenamefont {Frolov}, \citenamefont {Divinski}, \citenamefont {Wilde},\
  and\ \citenamefont {Liebscher}}]{ding_2024}%
  \BibitemOpen
  \bibfield  {author} {\bibinfo {author} {\bibfnamefont {H.}~\bibnamefont
  {Ding}}, \bibinfo {author} {\bibfnamefont {A.}~\bibnamefont {Akbari}},
  \bibinfo {author} {\bibfnamefont {E.}~\bibnamefont {Chen}}, \bibinfo {author}
  {\bibfnamefont {H.}~\bibnamefont {R{\"o}sner}}, \bibinfo {author}
  {\bibfnamefont {T.}~\bibnamefont {Frolov}}, \bibinfo {author} {\bibfnamefont
  {S.}~\bibnamefont {Divinski}}, \bibinfo {author} {\bibfnamefont
  {G.}~\bibnamefont {Wilde}},\ and\ \bibinfo {author} {\bibfnamefont {C.~H.}\
  \bibnamefont {Liebscher}},\ }\bibfield  {title} {\bibinfo {title} {Hierarchy
  of defects in near-{$\Sigma$5} tilt grain boundaries in copper studied by
  length-scale bridging electron microscopy},\ }\href
  {https://doi.org/10.1016/j.actamat.2025.120778} {\bibfield  {journal}
  {\bibinfo  {journal} {Acta Materialia}\ }\textbf {\bibinfo {volume} {287}},\
  \bibinfo {pages} {120778} (\bibinfo {year} {2025})}\BibitemShut {NoStop}%
\bibitem [{\citenamefont {Medlin}\ \emph {et~al.}(2017)\citenamefont {Medlin},
  \citenamefont {Hattar}, \citenamefont {Zimmerman}, \citenamefont
  {Abdeljawad},\ and\ \citenamefont {Foiles}}]{medlin_2017}%
  \BibitemOpen
  \bibfield  {author} {\bibinfo {author} {\bibfnamefont {D.~L.}\ \bibnamefont
  {Medlin}}, \bibinfo {author} {\bibfnamefont {K.}~\bibnamefont {Hattar}},
  \bibinfo {author} {\bibfnamefont {J.~A.}\ \bibnamefont {Zimmerman}}, \bibinfo
  {author} {\bibfnamefont {F.}~\bibnamefont {Abdeljawad}},\ and\ \bibinfo
  {author} {\bibfnamefont {S.~M.}\ \bibnamefont {Foiles}},\ }\bibfield  {title}
  {\bibinfo {title} {Defect character at grain boundary facet junctions:
  {Analysis} of an asymmetric {$\Sigma$}~=~5 grain boundary in {Fe}},\ }\href
  {https://doi.org/10.1016/j.actamat.2016.11.017} {\bibfield  {journal}
  {\bibinfo  {journal} {Acta Materialia}\ }\textbf {\bibinfo {volume} {124}},\
  \bibinfo {pages} {383} (\bibinfo {year} {2017})}\BibitemShut {NoStop}%
\bibitem [{\citenamefont {Zhou}\ \emph {et~al.}(2023)\citenamefont {Zhou},
  \citenamefont {Ahmadian}, \citenamefont {Gault}, \citenamefont {Ophus},
  \citenamefont {Liebscher}, \citenamefont {Dehm},\ and\ \citenamefont
  {Raabe}}]{zhou_2023}%
  \BibitemOpen
  \bibfield  {author} {\bibinfo {author} {\bibfnamefont {X.}~\bibnamefont
  {Zhou}}, \bibinfo {author} {\bibfnamefont {A.}~\bibnamefont {Ahmadian}},
  \bibinfo {author} {\bibfnamefont {B.}~\bibnamefont {Gault}}, \bibinfo
  {author} {\bibfnamefont {C.}~\bibnamefont {Ophus}}, \bibinfo {author}
  {\bibfnamefont {C.~H.}\ \bibnamefont {Liebscher}}, \bibinfo {author}
  {\bibfnamefont {G.}~\bibnamefont {Dehm}},\ and\ \bibinfo {author}
  {\bibfnamefont {D.}~\bibnamefont {Raabe}},\ }\bibfield  {title} {\bibinfo
  {title} {Atomic motifs govern the decoration of grain boundaries by
  interstitial solutes},\ }\href {https://doi.org/10.1038/s41467-023-39302-x}
  {\bibfield  {journal} {\bibinfo  {journal} {Nature Communications}\ }\textbf
  {\bibinfo {volume} {14}},\ \bibinfo {pages} {3535} (\bibinfo {year}
  {2023})}\BibitemShut {NoStop}%
\bibitem [{\citenamefont {Shamsuzzoha}\ \emph {et~al.}(1996)\citenamefont
  {Shamsuzzoha}, \citenamefont {Vazquez}, \citenamefont {Deymier},\ and\
  \citenamefont {Smith}}]{shamsuzzoha_1996}%
  \BibitemOpen
  \bibfield  {author} {\bibinfo {author} {\bibfnamefont {M.}~\bibnamefont
  {Shamsuzzoha}}, \bibinfo {author} {\bibfnamefont {I.}~\bibnamefont
  {Vazquez}}, \bibinfo {author} {\bibfnamefont {P.~A.}\ \bibnamefont
  {Deymier}},\ and\ \bibinfo {author} {\bibfnamefont {D.~J.}\ \bibnamefont
  {Smith}},\ }\bibfield  {title} {\bibinfo {title} {The atomic structure of a
  {$\Sigma$}=5[001]/(310) grain-boundary in an {Al}--5\% {Mg} alloy by
  high-resolution electron microscopy},\ }\href
  {https://doi.org/10.1007/BF00191050} {\bibfield  {journal} {\bibinfo
  {journal} {Interface Science}\ }\textbf {\bibinfo {volume} {3}},\ \bibinfo
  {pages} {227} (\bibinfo {year} {1996})}\BibitemShut {NoStop}%
\bibitem [{\citenamefont {Cosandey}\ \emph {et~al.}(1990)\citenamefont
  {Cosandey}, \citenamefont {Chan},\ and\ \citenamefont
  {Stadelmann}}]{cosandey_1990}%
  \BibitemOpen
  \bibfield  {author} {\bibinfo {author} {\bibfnamefont {F.}~\bibnamefont
  {Cosandey}}, \bibinfo {author} {\bibfnamefont {S.-W.}\ \bibnamefont {Chan}},\
  and\ \bibinfo {author} {\bibfnamefont {P.}~\bibnamefont {Stadelmann}},\
  }\bibfield  {title} {\bibinfo {title} {Atomic structure and energy of
  {$\Sigma$} = 5 tilt boundaries in gold},\ }\href
  {https://doi.org/10.1007/BF02646976} {\bibfield  {journal} {\bibinfo
  {journal} {Metallurgical Transactions A}\ }\textbf {\bibinfo {volume} {21}},\
  \bibinfo {pages} {2299} (\bibinfo {year} {1990})}\BibitemShut {NoStop}%
\bibitem [{\citenamefont {Campbell}\ \emph {et~al.}(1991)\citenamefont
  {Campbell}, \citenamefont {Foiles}, \citenamefont {King}, \citenamefont
  {R{\"u}hle},\ and\ \citenamefont {Wien}}]{campbell_1991}%
  \BibitemOpen
  \bibfield  {author} {\bibinfo {author} {\bibfnamefont {G.~H.}\ \bibnamefont
  {Campbell}}, \bibinfo {author} {\bibfnamefont {S.~M.}\ \bibnamefont
  {Foiles}}, \bibinfo {author} {\bibfnamefont {W.~E.}\ \bibnamefont {King}},
  \bibinfo {author} {\bibfnamefont {M.}~\bibnamefont {R{\"u}hle}},\ and\
  \bibinfo {author} {\bibfnamefont {W.}~\bibnamefont {Wien}},\ }\bibfield
  {title} {\bibinfo {title} {{HREM} investigation of the structure of the
  {$\Sigma$5}(210)/[001] symmetric tilt grain boundaries in {Nb}},\ }\href
  {https://doi.org/10.1557/PROC-229-191} {\bibfield  {journal} {\bibinfo
  {journal} {MRS Online Proceedings Library}\ }\textbf {\bibinfo {volume}
  {229}},\ \bibinfo {pages} {191} (\bibinfo {year} {1991})}\BibitemShut
  {NoStop}%
\bibitem [{\citenamefont {Campbell}\ \emph {et~al.}(1993)\citenamefont
  {Campbell}, \citenamefont {Foiles}, \citenamefont {Gumbsch}, \citenamefont
  {R{\"u}hle},\ and\ \citenamefont {King}}]{campbell_1993}%
  \BibitemOpen
  \bibfield  {author} {\bibinfo {author} {\bibfnamefont {G.~H.}\ \bibnamefont
  {Campbell}}, \bibinfo {author} {\bibfnamefont {S.~M.}\ \bibnamefont
  {Foiles}}, \bibinfo {author} {\bibfnamefont {P.}~\bibnamefont {Gumbsch}},
  \bibinfo {author} {\bibfnamefont {M.}~\bibnamefont {R{\"u}hle}},\ and\
  \bibinfo {author} {\bibfnamefont {W.~E.}\ \bibnamefont {King}},\ }\bibfield
  {title} {\bibinfo {title} {Atomic structure of the (310) twin in niobium:
  {Experimental} determination and comparison with theoretical predictions},\
  }\href {https://doi.org/10.1103/PhysRevLett.70.449} {\bibfield  {journal}
  {\bibinfo  {journal} {Physical Review Letters}\ }\textbf {\bibinfo {volume}
  {70}},\ \bibinfo {pages} {449} (\bibinfo {year} {1993})}\BibitemShut
  {NoStop}%
\bibitem [{\citenamefont {Campbell}\ \emph {et~al.}(2000)\citenamefont
  {Campbell}, \citenamefont {Belak},\ and\ \citenamefont
  {Moriarty}}]{campbell_2000}%
  \BibitemOpen
  \bibfield  {author} {\bibinfo {author} {\bibfnamefont {G.~H.}\ \bibnamefont
  {Campbell}}, \bibinfo {author} {\bibfnamefont {J.}~\bibnamefont {Belak}},\
  and\ \bibinfo {author} {\bibfnamefont {J.~A.}\ \bibnamefont {Moriarty}},\
  }\bibfield  {title} {\bibinfo {title} {Atomic structure of the {$\Sigma$}5
  (310)/[001] symmetric tilt grain boundary in tantalum},\ }\href
  {https://doi.org/10.1016/S1359-6462(00)00475-9} {\bibfield  {journal}
  {\bibinfo  {journal} {Scripta Materialia}\ }\textbf {\bibinfo {volume}
  {43}},\ \bibinfo {pages} {659} (\bibinfo {year} {2000})}\BibitemShut
  {NoStop}%
\bibitem [{\citenamefont {Campbell}\ \emph {et~al.}(1999)\citenamefont
  {Campbell}, \citenamefont {Belak},\ and\ \citenamefont
  {Moriarty}}]{campbell_1999}%
  \BibitemOpen
  \bibfield  {author} {\bibinfo {author} {\bibfnamefont {G.~H.}\ \bibnamefont
  {Campbell}}, \bibinfo {author} {\bibfnamefont {J.}~\bibnamefont {Belak}},\
  and\ \bibinfo {author} {\bibfnamefont {J.~A.}\ \bibnamefont {Moriarty}},\
  }\bibfield  {title} {\bibinfo {title} {Atomic structure of the {$\Sigma$5}
  (310)/[001] symmetric tilt grain boundary in molybdenum},\ }\href
  {https://doi.org/10.1016/S1359-6454(99)00258-X} {\bibfield  {journal}
  {\bibinfo  {journal} {Acta Materialia}\ }\textbf {\bibinfo {volume} {47}},\
  \bibinfo {pages} {3977} (\bibinfo {year} {1999})}\BibitemShut {NoStop}%
\bibitem [{\citenamefont {Bacia}\ \emph {et~al.}(1997)\citenamefont {Bacia},
  \citenamefont {Morillo}, \citenamefont {P{\'e}nisson},\ and\ \citenamefont
  {Pontikis}}]{bacia_1997}%
  \BibitemOpen
  \bibfield  {author} {\bibinfo {author} {\bibfnamefont {M.}~\bibnamefont
  {Bacia}}, \bibinfo {author} {\bibfnamefont {J.}~\bibnamefont {Morillo}},
  \bibinfo {author} {\bibfnamefont {J.~M.}\ \bibnamefont {P{\'e}nisson}},\ and\
  \bibinfo {author} {\bibfnamefont {V.}~\bibnamefont {Pontikis}},\ }\bibfield
  {title} {\bibinfo {title} {Atomic structure of the {$\Sigma$} = 5, (210) and
  (310), [001] tilt axis grain boundaries in {Mo}: a joint study by computer
  simulation and high-resolution electron microscopy},\ }\href
  {https://doi.org/10.1080/01418619708200009} {\bibfield  {journal} {\bibinfo
  {journal} {Philosophical Magazine A}\ }\textbf {\bibinfo {volume} {76}},\
  \bibinfo {pages} {945} (\bibinfo {year} {1997})}\BibitemShut {NoStop}%
\bibitem [{\citenamefont {Morita}\ and\ \citenamefont
  {Nakashima}(1997)}]{morita_1997}%
  \BibitemOpen
  \bibfield  {author} {\bibinfo {author} {\bibfnamefont {K.}~\bibnamefont
  {Morita}}\ and\ \bibinfo {author} {\bibfnamefont {H.}~\bibnamefont
  {Nakashima}},\ }\bibfield  {title} {\bibinfo {title} {Atomic periodicity of
  $\langle 001 \rangle$ symmetric tilt boundary in molybdenum},\ }\href
  {https://doi.org/10.1016/S0921-5093(97)00306-7} {\bibfield  {journal}
  {\bibinfo  {journal} {Materials Science and Engineering: A}\ }\textbf
  {\bibinfo {volume} {234--236}},\ \bibinfo {pages} {1053} (\bibinfo {year}
  {1997})}\BibitemShut {NoStop}%
\bibitem [{\citenamefont {Sutton}\ \emph {et~al.}(1983)\citenamefont {Sutton},
  \citenamefont {Vitek},\ and\ \citenamefont {Christian}}]{sutton_1983}%
  \BibitemOpen
  \bibfield  {author} {\bibinfo {author} {\bibfnamefont {A.~P.}\ \bibnamefont
  {Sutton}}, \bibinfo {author} {\bibfnamefont {V.}~\bibnamefont {Vitek}},\ and\
  \bibinfo {author} {\bibfnamefont {J.~W.}\ \bibnamefont {Christian}},\
  }\bibfield  {title} {\bibinfo {title} {On the structure of tilt grain
  boundaries in cubic metals {I}. symmetrical tilt boundaries},\ }\href
  {https://doi.org/10.1098/rsta.1983.0020} {\bibfield  {journal} {\bibinfo
  {journal} {Philosophical Transactions of the Royal Society of London. Series
  A}\ }\textbf {\bibinfo {volume} {309}},\ \bibinfo {pages} {1} (\bibinfo
  {year} {1983})}\BibitemShut {NoStop}%
\bibitem [{\citenamefont {Wolf}(1989)}]{wolf_1989}%
  \BibitemOpen
  \bibfield  {author} {\bibinfo {author} {\bibfnamefont {D.}~\bibnamefont
  {Wolf}},\ }\bibfield  {title} {\bibinfo {title} {Structure-energy correlation
  for grain boundaries in {F.C.C.} metals---{I}. boundaries on the (111) and
  (100) planes},\ }\href {https://doi.org/10.1016/0001-6160(89)90082-5}
  {\bibfield  {journal} {\bibinfo  {journal} {Acta Metallurgica}\ }\textbf
  {\bibinfo {volume} {37}},\ \bibinfo {pages} {1983} (\bibinfo {year}
  {1989})}\BibitemShut {NoStop}%
\bibitem [{\citenamefont {Ochs}\ \emph {et~al.}(2000)\citenamefont {Ochs},
  \citenamefont {Beck}, \citenamefont {Els{\"a}sser},\ and\ \citenamefont
  {Meyer}}]{ochs_2000a}%
  \BibitemOpen
  \bibfield  {author} {\bibinfo {author} {\bibfnamefont {T.}~\bibnamefont
  {Ochs}}, \bibinfo {author} {\bibfnamefont {O.}~\bibnamefont {Beck}}, \bibinfo
  {author} {\bibfnamefont {C.}~\bibnamefont {Els{\"a}sser}},\ and\ \bibinfo
  {author} {\bibfnamefont {B.}~\bibnamefont {Meyer}},\ }\bibfield  {title}
  {\bibinfo {title} {Symmetrical tilt grain boundaries in body-centred cubic
  transition metals: {An} \emph{ab initio} local-density-functional study},\
  }\href {https://doi.org/10.1080/01418610008212057} {\bibfield  {journal}
  {\bibinfo  {journal} {Philosophical Magazine A}\ }\textbf {\bibinfo {volume}
  {80}},\ \bibinfo {pages} {351} (\bibinfo {year} {2000})}\BibitemShut
  {NoStop}%
\bibitem [{\citenamefont {Scheiber}\ \emph {et~al.}(2016)\citenamefont
  {Scheiber}, \citenamefont {Pippan}, \citenamefont {Puschnig},\ and\
  \citenamefont {Romaner}}]{scheiber_2016}%
  \BibitemOpen
  \bibfield  {author} {\bibinfo {author} {\bibfnamefont {D.}~\bibnamefont
  {Scheiber}}, \bibinfo {author} {\bibfnamefont {R.}~\bibnamefont {Pippan}},
  \bibinfo {author} {\bibfnamefont {P.}~\bibnamefont {Puschnig}},\ and\
  \bibinfo {author} {\bibfnamefont {L.}~\bibnamefont {Romaner}},\ }\bibfield
  {title} {\bibinfo {title} {\emph{Ab initio} calculations of grain boundaries
  in bcc metals},\ }\href {https://doi.org/10.1088/0965-0393/24/3/035013}
  {\bibfield  {journal} {\bibinfo  {journal} {Modelling and Simulation in
  Materials Science and Engineering}\ }\textbf {\bibinfo {volume} {24}},\
  \bibinfo {pages} {035013} (\bibinfo {year} {2016})}\BibitemShut {NoStop}%
\bibitem [{\citenamefont {Zheng}\ \emph {et~al.}(2020)\citenamefont {Zheng},
  \citenamefont {Li}, \citenamefont {Tran}, \citenamefont {Chen}, \citenamefont
  {Horton}, \citenamefont {Winston}, \citenamefont {Persson},\ and\
  \citenamefont {Ong}}]{zheng_2020}%
  \BibitemOpen
  \bibfield  {author} {\bibinfo {author} {\bibfnamefont {H.}~\bibnamefont
  {Zheng}}, \bibinfo {author} {\bibfnamefont {X.-G.}\ \bibnamefont {Li}},
  \bibinfo {author} {\bibfnamefont {R.}~\bibnamefont {Tran}}, \bibinfo {author}
  {\bibfnamefont {C.}~\bibnamefont {Chen}}, \bibinfo {author} {\bibfnamefont
  {M.}~\bibnamefont {Horton}}, \bibinfo {author} {\bibfnamefont
  {D.}~\bibnamefont {Winston}}, \bibinfo {author} {\bibfnamefont {K.~A.}\
  \bibnamefont {Persson}},\ and\ \bibinfo {author} {\bibfnamefont {S.~P.}\
  \bibnamefont {Ong}},\ }\bibfield  {title} {\bibinfo {title} {Grain boundary
  properties of elemental metals},\ }\href
  {https://doi.org/10.1016/j.actamat.2019.12.030} {\bibfield  {journal}
  {\bibinfo  {journal} {Acta Materialia}\ }\textbf {\bibinfo {volume} {186}},\
  \bibinfo {pages} {40} (\bibinfo {year} {2020})}\BibitemShut {NoStop}%
\bibitem [{\citenamefont {Mishin}\ and\ \citenamefont
  {Farkas}(1998)}]{mishin_1998}%
  \BibitemOpen
  \bibfield  {author} {\bibinfo {author} {\bibfnamefont {Y.}~\bibnamefont
  {Mishin}}\ and\ \bibinfo {author} {\bibfnamefont {D.}~\bibnamefont
  {Farkas}},\ }\bibfield  {title} {\bibinfo {title} {Atomistic simulation of
  [001] symmetrical tilt grain boundaries in {NiAl}},\ }\href
  {https://doi.org/10.1080/014186198253679} {\bibfield  {journal} {\bibinfo
  {journal} {Philosophical Magazine A}\ }\textbf {\bibinfo {volume} {78}},\
  \bibinfo {pages} {29} (\bibinfo {year} {1998})}\BibitemShut {NoStop}%
\bibitem [{\citenamefont {Frolov}\ \emph {et~al.}(2013)\citenamefont {Frolov},
  \citenamefont {Olmsted}, \citenamefont {Asta},\ and\ \citenamefont
  {Mishin}}]{frolov_2013_gbphase}%
  \BibitemOpen
  \bibfield  {author} {\bibinfo {author} {\bibfnamefont {T.}~\bibnamefont
  {Frolov}}, \bibinfo {author} {\bibfnamefont {D.~L.}\ \bibnamefont {Olmsted}},
  \bibinfo {author} {\bibfnamefont {M.}~\bibnamefont {Asta}},\ and\ \bibinfo
  {author} {\bibfnamefont {Y.}~\bibnamefont {Mishin}},\ }\bibfield  {title}
  {\bibinfo {title} {Structural phase transformations in metallic grain
  boundaries},\ }\href {https://doi.org/10.1038/ncomms2919} {\bibfield
  {journal} {\bibinfo  {journal} {Nature Communications}\ }\textbf {\bibinfo
  {volume} {4}},\ \bibinfo {pages} {1} (\bibinfo {year} {2013})}\BibitemShut
  {NoStop}%
\bibitem [{\citenamefont {Zhu}\ \emph {et~al.}(2018)\citenamefont {Zhu},
  \citenamefont {Samanta}, \citenamefont {Li}, \citenamefont {Rudd},\ and\
  \citenamefont {Frolov}}]{zhu_2018}%
  \BibitemOpen
  \bibfield  {author} {\bibinfo {author} {\bibfnamefont {Q.}~\bibnamefont
  {Zhu}}, \bibinfo {author} {\bibfnamefont {A.}~\bibnamefont {Samanta}},
  \bibinfo {author} {\bibfnamefont {B.}~\bibnamefont {Li}}, \bibinfo {author}
  {\bibfnamefont {R.~E.}\ \bibnamefont {Rudd}},\ and\ \bibinfo {author}
  {\bibfnamefont {T.}~\bibnamefont {Frolov}},\ }\bibfield  {title} {\bibinfo
  {title} {Predicting phase behavior of grain boundaries with evolutionary
  search and machine learning},\ }\href
  {https://doi.org/10.1038/s41467-018-02937-2} {\bibfield  {journal} {\bibinfo
  {journal} {Nature Communications}\ }\textbf {\bibinfo {volume} {9}},\
  \bibinfo {pages} {467} (\bibinfo {year} {2018})}\BibitemShut {NoStop}%
\bibitem [{\citenamefont {Frolov}\ \emph
  {et~al.}(2018{\natexlab{a}})\citenamefont {Frolov}, \citenamefont {Setyawan},
  \citenamefont {Kurtz}, \citenamefont {Marian}, \citenamefont {Oganov},
  \citenamefont {Rudd},\ and\ \citenamefont {Zhu}}]{frolov_2018}%
  \BibitemOpen
  \bibfield  {author} {\bibinfo {author} {\bibfnamefont {T.}~\bibnamefont
  {Frolov}}, \bibinfo {author} {\bibfnamefont {W.}~\bibnamefont {Setyawan}},
  \bibinfo {author} {\bibfnamefont {R.~J.}\ \bibnamefont {Kurtz}}, \bibinfo
  {author} {\bibfnamefont {J.}~\bibnamefont {Marian}}, \bibinfo {author}
  {\bibfnamefont {A.~E.}\ \bibnamefont {Oganov}}, \bibinfo {author}
  {\bibfnamefont {R.~E.}\ \bibnamefont {Rudd}},\ and\ \bibinfo {author}
  {\bibfnamefont {Q.}~\bibnamefont {Zhu}},\ }\bibfield  {title} {\bibinfo
  {title} {Grain boundary phases in bcc metals},\ }\href
  {https://doi.org/10.1039/C8NR00271A} {\bibfield  {journal} {\bibinfo
  {journal} {Nanoscale}\ }\textbf {\bibinfo {volume} {10}},\ \bibinfo {pages}
  {8253} (\bibinfo {year} {2018}{\natexlab{a}})}\BibitemShut {NoStop}%
\bibitem [{\citenamefont {Seki}\ \emph {et~al.}(2023)\citenamefont {Seki},
  \citenamefont {Futazuka}, \citenamefont {Morishige}, \citenamefont
  {Matsubara}, \citenamefont {Ikuhara},\ and\ \citenamefont
  {Shibata}}]{seki_2023}%
  \BibitemOpen
  \bibfield  {author} {\bibinfo {author} {\bibfnamefont {T.}~\bibnamefont
  {Seki}}, \bibinfo {author} {\bibfnamefont {T.}~\bibnamefont {Futazuka}},
  \bibinfo {author} {\bibfnamefont {N.}~\bibnamefont {Morishige}}, \bibinfo
  {author} {\bibfnamefont {R.}~\bibnamefont {Matsubara}}, \bibinfo {author}
  {\bibfnamefont {Y.}~\bibnamefont {Ikuhara}},\ and\ \bibinfo {author}
  {\bibfnamefont {N.}~\bibnamefont {Shibata}},\ }\bibfield  {title} {\bibinfo
  {title} {Incommensurate grain-boundary atomic structure},\ }\href
  {https://doi.org/10.1038/s41467-023-43536-0} {\bibfield  {journal} {\bibinfo
  {journal} {Nature Communications}\ }\textbf {\bibinfo {volume} {14}},\
  \bibinfo {pages} {7806} (\bibinfo {year} {2023})}\BibitemShut {NoStop}%
\bibitem [{\citenamefont {Pemma}\ \emph {et~al.}(2024)\citenamefont {Pemma},
  \citenamefont {Janisch}, \citenamefont {Dehm},\ and\ \citenamefont
  {Brink}}]{pemma_2024}%
  \BibitemOpen
  \bibfield  {author} {\bibinfo {author} {\bibfnamefont {S.}~\bibnamefont
  {Pemma}}, \bibinfo {author} {\bibfnamefont {R.}~\bibnamefont {Janisch}},
  \bibinfo {author} {\bibfnamefont {G.}~\bibnamefont {Dehm}},\ and\ \bibinfo
  {author} {\bibfnamefont {T.}~\bibnamefont {Brink}},\ }\bibfield  {title}
  {\bibinfo {title} {Effect of the atomic structure of complexions on the
  active disconnection mode during shear-coupled grain boundary motion},\
  }\href {https://doi.org/10.1103/PhysRevMaterials.8.063602} {\bibfield
  {journal} {\bibinfo  {journal} {Physical Review Materials}\ }\textbf
  {\bibinfo {volume} {8}},\ \bibinfo {pages} {063602} (\bibinfo {year}
  {2024})}\BibitemShut {NoStop}%
\bibitem [{\citenamefont {Rajeshwari~K.}\ \emph {et~al.}(2020)\citenamefont
  {Rajeshwari~K.}, \citenamefont {Sankaran}, \citenamefont {Hari~Kumar},
  \citenamefont {R{\"o}sner}, \citenamefont {Peterlechner}, \citenamefont
  {Esin}, \citenamefont {Divinski},\ and\ \citenamefont
  {Wilde}}]{rajeshwari_2020}%
  \BibitemOpen
  \bibfield  {author} {\bibinfo {author} {\bibfnamefont {S.}~\bibnamefont
  {Rajeshwari~K.}}, \bibinfo {author} {\bibfnamefont {S.}~\bibnamefont
  {Sankaran}}, \bibinfo {author} {\bibfnamefont {K.~C.}\ \bibnamefont
  {Hari~Kumar}}, \bibinfo {author} {\bibfnamefont {H.}~\bibnamefont
  {R{\"o}sner}}, \bibinfo {author} {\bibfnamefont {M.}~\bibnamefont
  {Peterlechner}}, \bibinfo {author} {\bibfnamefont {V.~A.}\ \bibnamefont
  {Esin}}, \bibinfo {author} {\bibfnamefont {S.}~\bibnamefont {Divinski}},\
  and\ \bibinfo {author} {\bibfnamefont {G.}~\bibnamefont {Wilde}},\ }\bibfield
   {title} {\bibinfo {title} {Grain boundary diffusion and grain boundary
  structures of a {Ni-Cr-Fe-} alloy: {Evidences} for grain boundary phase
  transitions},\ }\href {https://doi.org/10.1016/j.actamat.2020.05.051}
  {\bibfield  {journal} {\bibinfo  {journal} {Acta Materialia}\ }\textbf
  {\bibinfo {volume} {195}},\ \bibinfo {pages} {501} (\bibinfo {year}
  {2020})}\BibitemShut {NoStop}%
\bibitem [{\citenamefont {Chen}\ \emph {et~al.}(2024)\citenamefont {Chen},
  \citenamefont {Heo}, \citenamefont {Wood}, \citenamefont {Asta},\ and\
  \citenamefont {Frolov}}]{chen_2024}%
  \BibitemOpen
  \bibfield  {author} {\bibinfo {author} {\bibfnamefont {E.}~\bibnamefont
  {Chen}}, \bibinfo {author} {\bibfnamefont {T.~W.}\ \bibnamefont {Heo}},
  \bibinfo {author} {\bibfnamefont {B.~C.}\ \bibnamefont {Wood}}, \bibinfo
  {author} {\bibfnamefont {M.}~\bibnamefont {Asta}},\ and\ \bibinfo {author}
  {\bibfnamefont {T.}~\bibnamefont {Frolov}},\ }\bibfield  {title} {\bibinfo
  {title} {Grand canonically optimized grain boundary phases in hexagonal
  close-packed titanium},\ }\href {https://doi.org/10.1038/s41467-024-51330-9}
  {\bibfield  {journal} {\bibinfo  {journal} {Nature Communications}\ }\textbf
  {\bibinfo {volume} {15}},\ \bibinfo {pages} {7049} (\bibinfo {year}
  {2024})}\BibitemShut {NoStop}%
\bibitem [{sup()}]{supp}%
  \BibitemOpen
  \href@noop {} {}\bibinfo {note} {See Supplemental Material at [URL will be
  inserted by publisher] for additional grain boundary structures and
  methodology details.}\BibitemShut {Stop}%
\bibitem [{\citenamefont {Chesser}\ \emph {et~al.}(2024)\citenamefont
  {Chesser}, \citenamefont {Koju},\ and\ \citenamefont
  {Mishin}}]{chesser_2024}%
  \BibitemOpen
  \bibfield  {author} {\bibinfo {author} {\bibfnamefont {I.}~\bibnamefont
  {Chesser}}, \bibinfo {author} {\bibfnamefont {R.~K.}\ \bibnamefont {Koju}},\
  and\ \bibinfo {author} {\bibfnamefont {Y.}~\bibnamefont {Mishin}},\
  }\bibfield  {title} {\bibinfo {title} {Atomic-level mechanisms of
  short-circuit diffusion in materials},\ }\href
  {https://doi.org/10.1515/ijmr-2023-0202} {\bibfield  {journal} {\bibinfo
  {journal} {International Journal of Materials Research}\ }\textbf {\bibinfo
  {volume} {115}},\ \bibinfo {pages} {85} (\bibinfo {year} {2024})}\BibitemShut
  {NoStop}%
\bibitem [{\citenamefont {Mahmood}\ \emph {et~al.}(2024)\citenamefont
  {Mahmood}, \citenamefont {Daw}, \citenamefont {Chandross},\ and\
  \citenamefont {Abdeljawad}}]{mahmood_2024}%
  \BibitemOpen
  \bibfield  {author} {\bibinfo {author} {\bibfnamefont {Y.}~\bibnamefont
  {Mahmood}}, \bibinfo {author} {\bibfnamefont {M.~S.}\ \bibnamefont {Daw}},
  \bibinfo {author} {\bibfnamefont {M.}~\bibnamefont {Chandross}},\ and\
  \bibinfo {author} {\bibfnamefont {F.}~\bibnamefont {Abdeljawad}},\ }\bibfield
   {title} {\bibinfo {title} {Universal trends in computed grain boundary
  energies of {FCC} metals},\ }\href
  {https://doi.org/10.1016/j.scriptamat.2023.115900} {\bibfield  {journal}
  {\bibinfo  {journal} {Scripta Materialia}\ }\textbf {\bibinfo {volume}
  {242}},\ \bibinfo {pages} {115900} (\bibinfo {year} {2024})}\BibitemShut
  {NoStop}%
\bibitem [{\citenamefont {Freitas}\ \emph {et~al.}(2018)\citenamefont
  {Freitas}, \citenamefont {Rudd}, \citenamefont {Asta},\ and\ \citenamefont
  {Frolov}}]{freitas_2018}%
  \BibitemOpen
  \bibfield  {author} {\bibinfo {author} {\bibfnamefont {R.}~\bibnamefont
  {Freitas}}, \bibinfo {author} {\bibfnamefont {R.~E.}\ \bibnamefont {Rudd}},
  \bibinfo {author} {\bibfnamefont {M.}~\bibnamefont {Asta}},\ and\ \bibinfo
  {author} {\bibfnamefont {T.}~\bibnamefont {Frolov}},\ }\bibfield  {title}
  {\bibinfo {title} {Free energy of grain boundary phases: Atomistic
  calculations for {$\Sigma$5}(310)[001] grain boundary in {Cu}},\ }\href
  {https://doi.org/10.1103/PhysRevMaterials.2.093603} {\bibfield  {journal}
  {\bibinfo  {journal} {Physical Review Materials}\ }\textbf {\bibinfo {volume}
  {2}},\ \bibinfo {pages} {093603} (\bibinfo {year} {2018})}\BibitemShut
  {NoStop}%
\bibitem [{\citenamefont {Madsen}\ and\ \citenamefont
  {Susi}(2021)}]{madsen_2021}%
  \BibitemOpen
  \bibfield  {author} {\bibinfo {author} {\bibfnamefont {J.}~\bibnamefont
  {Madsen}}\ and\ \bibinfo {author} {\bibfnamefont {T.}~\bibnamefont {Susi}},\
  }\bibfield  {title} {\bibinfo {title} {The {abTEM} code: transmission
  electron microscopy from first principles},\ }\bibfield  {journal} {\bibinfo
  {journal} {Open Research Europe}\ }\textbf {\bibinfo {volume} {1}},\ \href
  {https://doi.org/10.12688/openreseurope.13015.2}
  {10.12688/openreseurope.13015.2} (\bibinfo {year} {2021})\BibitemShut
  {NoStop}%
\bibitem [{\citenamefont {Campbell}\ \emph {et~al.}(2002)\citenamefont
  {Campbell}, \citenamefont {Kumar}, \citenamefont {King}, \citenamefont
  {Belak}, \citenamefont {Moriarty},\ and\ \citenamefont
  {Foiles}}]{campbell_2002}%
  \BibitemOpen
  \bibfield  {author} {\bibinfo {author} {\bibfnamefont {G.~H.}\ \bibnamefont
  {Campbell}}, \bibinfo {author} {\bibfnamefont {M.}~\bibnamefont {Kumar}},
  \bibinfo {author} {\bibfnamefont {W.~E.}\ \bibnamefont {King}}, \bibinfo
  {author} {\bibfnamefont {J.}~\bibnamefont {Belak}}, \bibinfo {author}
  {\bibfnamefont {J.~A.}\ \bibnamefont {Moriarty}},\ and\ \bibinfo {author}
  {\bibfnamefont {S.~M.}\ \bibnamefont {Foiles}},\ }\bibfield  {title}
  {\bibinfo {title} {The rigid-body displacement observed at the {$\Sigma$} =
  5, (310)-[001] symmetric tilt grain boundary in central transition bcc
  metals},\ }\href {https://doi.org/10.1080/01418610208240038} {\bibfield
  {journal} {\bibinfo  {journal} {Philosophical Magazine A}\ }\textbf {\bibinfo
  {volume} {82}},\ \bibinfo {pages} {1573} (\bibinfo {year}
  {2002})}\BibitemShut {NoStop}%
\bibitem [{\citenamefont {Hunter}(2007)}]{hunter_2007}%
  \BibitemOpen
  \bibfield  {author} {\bibinfo {author} {\bibfnamefont {J.~D.}\ \bibnamefont
  {Hunter}},\ }\bibfield  {title} {\bibinfo {title} {Matplotlib: A {2D}
  graphics environment},\ }\href {https://doi.org/10.1109/MCSE.2007.55}
  {\bibfield  {journal} {\bibinfo  {journal} {Computing in Science \&
  Engineering}\ }\textbf {\bibinfo {volume} {9}},\ \bibinfo {pages} {90}
  (\bibinfo {year} {2007})}\BibitemShut {NoStop}%
\bibitem [{\citenamefont {Stukowski}(2009)}]{stukowski_2009}%
  \BibitemOpen
  \bibfield  {author} {\bibinfo {author} {\bibfnamefont {A.}~\bibnamefont
  {Stukowski}},\ }\bibfield  {title} {\bibinfo {title} {Visualization and
  analysis of atomistic simulation data with {OVITO}---the {Open Visualization
  Tool}},\ }\href {https://doi.org/10.1088/0965-0393/18/1/015012} {\bibfield
  {journal} {\bibinfo  {journal} {Modelling and Simulation in Materials Science
  and Engineering}\ }\textbf {\bibinfo {volume} {18}},\ \bibinfo {pages}
  {015012} (\bibinfo {year} {2009})}\BibitemShut {NoStop}%
\bibitem [{\citenamefont {Chen}\ and\ \citenamefont {Frolov}(2025)}]{zenodo}%
  \BibitemOpen
  \bibfield  {author} {\bibinfo {author} {\bibfnamefont {E.}~\bibnamefont
  {Chen}}\ and\ \bibinfo {author} {\bibfnamefont {T.}~\bibnamefont {Frolov}},\
  }\href@noop {} {\bibinfo {title} {Quasi-aperiodic grain boundary phases of
  {$\Sigma5$} tilt grain boundaries in refractory metals}},\ \bibinfo
  {howpublished} {\emph{Zenodo} v1
  \url{https://doi.org/10.5281/zenodo.14737236}} (\bibinfo {year}
  {2025})\BibitemShut {NoStop}%
\bibitem [{\citenamefont {Daw}\ and\ \citenamefont {Baskes}(1984)}]{daw_1984}%
  \BibitemOpen
  \bibfield  {author} {\bibinfo {author} {\bibfnamefont {M.~S.}\ \bibnamefont
  {Daw}}\ and\ \bibinfo {author} {\bibfnamefont {M.~I.}\ \bibnamefont
  {Baskes}},\ }\bibfield  {title} {\bibinfo {title} {Embedded-atom method:
  Derivation and application to impurities, surfaces, and other defects in
  metals},\ }\href {https://doi.org/10.1103/PhysRevB.29.6443} {\bibfield
  {journal} {\bibinfo  {journal} {Physical Review B}\ }\textbf {\bibinfo
  {volume} {29}},\ \bibinfo {pages} {6443} (\bibinfo {year}
  {1984})}\BibitemShut {NoStop}%
\bibitem [{\citenamefont {Baskes}(1992)}]{baskes_1992}%
  \BibitemOpen
  \bibfield  {author} {\bibinfo {author} {\bibfnamefont {M.~I.}\ \bibnamefont
  {Baskes}},\ }\bibfield  {title} {\bibinfo {title} {Modified embedded-atom
  potentials for cubic materials and impurities},\ }\href
  {https://doi.org/10.1103/PhysRevB.46.2727} {\bibfield  {journal} {\bibinfo
  {journal} {Physical Review B}\ }\textbf {\bibinfo {volume} {46}},\ \bibinfo
  {pages} {2727} (\bibinfo {year} {1992})}\BibitemShut {NoStop}%
\bibitem [{\citenamefont {Mishin}\ \emph {et~al.}(2005)\citenamefont {Mishin},
  \citenamefont {Mehl},\ and\ \citenamefont
  {Papaconstantopoulos}}]{mishin_2005}%
  \BibitemOpen
  \bibfield  {author} {\bibinfo {author} {\bibfnamefont {Y.}~\bibnamefont
  {Mishin}}, \bibinfo {author} {\bibfnamefont {M.~J.}\ \bibnamefont {Mehl}},\
  and\ \bibinfo {author} {\bibfnamefont {D.~A.}\ \bibnamefont
  {Papaconstantopoulos}},\ }\bibfield  {title} {\bibinfo {title} {Phase
  stability in the {Fe}--{Ni} system: Investigation by first-principles
  calculations and atomistic simulations},\ }\href
  {https://doi.org/10.1016/j.actamat.2005.05.001} {\bibfield  {journal}
  {\bibinfo  {journal} {Acta Materialia}\ }\textbf {\bibinfo {volume} {53}},\
  \bibinfo {pages} {4029} (\bibinfo {year} {2005})}\BibitemShut {NoStop}%
\bibitem [{\citenamefont {Fellinger}\ \emph {et~al.}(2010)\citenamefont
  {Fellinger}, \citenamefont {Park},\ and\ \citenamefont
  {Wilkins}}]{fellinger_2010}%
  \BibitemOpen
  \bibfield  {author} {\bibinfo {author} {\bibfnamefont {M.~R.}\ \bibnamefont
  {Fellinger}}, \bibinfo {author} {\bibfnamefont {H.}~\bibnamefont {Park}},\
  and\ \bibinfo {author} {\bibfnamefont {J.~W.}\ \bibnamefont {Wilkins}},\
  }\bibfield  {title} {\bibinfo {title} {Force-matched embedded-atom method
  potential for niobium},\ }\href {https://doi.org/10.1103/PhysRevB.81.144119}
  {\bibfield  {journal} {\bibinfo  {journal} {Physical Review B}\ }\textbf
  {\bibinfo {volume} {81}},\ \bibinfo {pages} {144119} (\bibinfo {year}
  {2010})}\BibitemShut {NoStop}%
\bibitem [{\citenamefont {Li}\ \emph {et~al.}(2003)\citenamefont {Li},
  \citenamefont {Siegel}, \citenamefont {Adams},\ and\ \citenamefont
  {Liu}}]{li_2003}%
  \BibitemOpen
  \bibfield  {author} {\bibinfo {author} {\bibfnamefont {Y.}~\bibnamefont
  {Li}}, \bibinfo {author} {\bibfnamefont {D.~J.}\ \bibnamefont {Siegel}},
  \bibinfo {author} {\bibfnamefont {J.~B.}\ \bibnamefont {Adams}},\ and\
  \bibinfo {author} {\bibfnamefont {X.-Y.}\ \bibnamefont {Liu}},\ }\bibfield
  {title} {\bibinfo {title} {Embedded-atom-method tantalum potential developed
  by the force-matching method},\ }\href
  {https://doi.org/10.1103/PhysRevB.67.125101} {\bibfield  {journal} {\bibinfo
  {journal} {Physical Review B}\ }\textbf {\bibinfo {volume} {67}},\ \bibinfo
  {pages} {125101} (\bibinfo {year} {2003})}\BibitemShut {NoStop}%
\bibitem [{\citenamefont {Zhou}\ \emph {et~al.}(2004)\citenamefont {Zhou},
  \citenamefont {Johnson},\ and\ \citenamefont {Wadley}}]{zhou_2004}%
  \BibitemOpen
  \bibfield  {author} {\bibinfo {author} {\bibfnamefont {X.}~\bibnamefont
  {Zhou}}, \bibinfo {author} {\bibfnamefont {R.~A.}\ \bibnamefont {Johnson}},\
  and\ \bibinfo {author} {\bibfnamefont {H.~N.~G.}\ \bibnamefont {Wadley}},\
  }\bibfield  {title} {\bibinfo {title} {Misfit-energy-increasing dislocations
  in vapor-deposited {CoFe}/{NiFe} multilayers},\ }\href
  {https://doi.org/10.1103/PhysRevB.69.144113} {\bibfield  {journal} {\bibinfo
  {journal} {Physical Review B}\ }\textbf {\bibinfo {volume} {69}},\ \bibinfo
  {pages} {144113} (\bibinfo {year} {2004})}\BibitemShut {NoStop}%
\bibitem [{\citenamefont {Marinica}\ \emph {et~al.}(2013)\citenamefont
  {Marinica}, \citenamefont {Ventelon}, \citenamefont {Gilbert}, \citenamefont
  {Proville}, \citenamefont {Dudarev}, \citenamefont {Marian}, \citenamefont
  {Bencteux},\ and\ \citenamefont {Willaime}}]{marinica_2013}%
  \BibitemOpen
  \bibfield  {author} {\bibinfo {author} {\bibfnamefont {M.-C.}\ \bibnamefont
  {Marinica}}, \bibinfo {author} {\bibfnamefont {L.}~\bibnamefont {Ventelon}},
  \bibinfo {author} {\bibfnamefont {M.~R.}\ \bibnamefont {Gilbert}}, \bibinfo
  {author} {\bibfnamefont {L.}~\bibnamefont {Proville}}, \bibinfo {author}
  {\bibfnamefont {S.~L.}\ \bibnamefont {Dudarev}}, \bibinfo {author}
  {\bibfnamefont {J.}~\bibnamefont {Marian}}, \bibinfo {author} {\bibfnamefont
  {G.}~\bibnamefont {Bencteux}},\ and\ \bibinfo {author} {\bibfnamefont
  {F.}~\bibnamefont {Willaime}},\ }\bibfield  {title} {\bibinfo {title}
  {Interatomic potentials for modelling radiation defects and dislocations in
  tungsten},\ }\href {https://doi.org/10.1088/0953-8984/25/39/395502}
  {\bibfield  {journal} {\bibinfo  {journal} {Journal of Physics: Condensed
  Matter}\ }\textbf {\bibinfo {volume} {25}},\ \bibinfo {pages} {395502}
  (\bibinfo {year} {2013})}\BibitemShut {NoStop}%
\bibitem [{\citenamefont {Proville}\ \emph {et~al.}(2012)\citenamefont
  {Proville}, \citenamefont {Rodney},\ and\ \citenamefont
  {Marinica}}]{proville_2012}%
  \BibitemOpen
  \bibfield  {author} {\bibinfo {author} {\bibfnamefont {L.}~\bibnamefont
  {Proville}}, \bibinfo {author} {\bibfnamefont {D.}~\bibnamefont {Rodney}},\
  and\ \bibinfo {author} {\bibfnamefont {M.-C.}\ \bibnamefont {Marinica}},\
  }\bibfield  {title} {\bibinfo {title} {Quantum effect on thermally activated
  glide of dislocations},\ }\href {https://doi.org/10.1038/nmat3401} {\bibfield
   {journal} {\bibinfo  {journal} {Nature Materials}\ }\textbf {\bibinfo
  {volume} {11}},\ \bibinfo {pages} {845} (\bibinfo {year} {2012})}\BibitemShut
  {NoStop}%
\bibitem [{\citenamefont {Yang}\ and\ \citenamefont {Qi}(2019)}]{yang_2019}%
  \BibitemOpen
  \bibfield  {author} {\bibinfo {author} {\bibfnamefont {C.}~\bibnamefont
  {Yang}}\ and\ \bibinfo {author} {\bibfnamefont {L.}~\bibnamefont {Qi}},\
  }\bibfield  {title} {\bibinfo {title} {Modified embedded-atom method
  potential of niobium for studies on mechanical properties},\ }\href
  {https://doi.org/10.1016/j.commatsci.2019.01.047} {\bibfield  {journal}
  {\bibinfo  {journal} {Computational Materials Science}\ }\textbf {\bibinfo
  {volume} {161}},\ \bibinfo {pages} {351} (\bibinfo {year}
  {2019})}\BibitemShut {NoStop}%
\bibitem [{\citenamefont {Park}\ \emph {et~al.}(2012)\citenamefont {Park},
  \citenamefont {Fellinger}, \citenamefont {Lenosky}, \citenamefont {Tipton},
  \citenamefont {Trinkle}, \citenamefont {Rudin}, \citenamefont {Woodward},
  \citenamefont {Wilkins},\ and\ \citenamefont {Hennig}}]{park_2012}%
  \BibitemOpen
  \bibfield  {author} {\bibinfo {author} {\bibfnamefont {H.}~\bibnamefont
  {Park}}, \bibinfo {author} {\bibfnamefont {M.~R.}\ \bibnamefont {Fellinger}},
  \bibinfo {author} {\bibfnamefont {T.~J.}\ \bibnamefont {Lenosky}}, \bibinfo
  {author} {\bibfnamefont {W.~W.}\ \bibnamefont {Tipton}}, \bibinfo {author}
  {\bibfnamefont {D.~R.}\ \bibnamefont {Trinkle}}, \bibinfo {author}
  {\bibfnamefont {S.~P.}\ \bibnamefont {Rudin}}, \bibinfo {author}
  {\bibfnamefont {C.}~\bibnamefont {Woodward}}, \bibinfo {author}
  {\bibfnamefont {J.~W.}\ \bibnamefont {Wilkins}},\ and\ \bibinfo {author}
  {\bibfnamefont {R.~G.}\ \bibnamefont {Hennig}},\ }\bibfield  {title}
  {\bibinfo {title} {\emph{Ab initio} based empirical potential used to study
  the mechanical properties of molybdenum},\ }\href
  {https://doi.org/10.1103/PhysRevB.85.214121} {\bibfield  {journal} {\bibinfo
  {journal} {Physical Review B}\ }\textbf {\bibinfo {volume} {85}},\ \bibinfo
  {pages} {214121} (\bibinfo {year} {2012})}\BibitemShut {NoStop}%
\bibitem [{\citenamefont {Hiremath}\ \emph {et~al.}(2022)\citenamefont
  {Hiremath}, \citenamefont {Melin}, \citenamefont {Bitzek},\ and\
  \citenamefont {Olsson}}]{hiremath_2022}%
  \BibitemOpen
  \bibfield  {author} {\bibinfo {author} {\bibfnamefont {P.}~\bibnamefont
  {Hiremath}}, \bibinfo {author} {\bibfnamefont {S.}~\bibnamefont {Melin}},
  \bibinfo {author} {\bibfnamefont {E.}~\bibnamefont {Bitzek}},\ and\ \bibinfo
  {author} {\bibfnamefont {P.~A.~T.}\ \bibnamefont {Olsson}},\ }\bibfield
  {title} {\bibinfo {title} {Effects of interatomic potential on fracture
  behaviour in single- and bicrystalline tungsten},\ }\href
  {https://doi.org/10.1016/j.commatsci.2022.111283} {\bibfield  {journal}
  {\bibinfo  {journal} {Computational Materials Science}\ }\textbf {\bibinfo
  {volume} {207}},\ \bibinfo {pages} {111283} (\bibinfo {year}
  {2022})}\BibitemShut {NoStop}%
\bibitem [{\citenamefont {Purja~Pun}\ \emph {et~al.}(2015)\citenamefont
  {Purja~Pun}, \citenamefont {Darling}, \citenamefont {Kecskes},\ and\
  \citenamefont {Mishin}}]{purjapun_2015}%
  \BibitemOpen
  \bibfield  {author} {\bibinfo {author} {\bibfnamefont {G.~P.}\ \bibnamefont
  {Purja~Pun}}, \bibinfo {author} {\bibfnamefont {K.~A.}\ \bibnamefont
  {Darling}}, \bibinfo {author} {\bibfnamefont {L.~J.}\ \bibnamefont
  {Kecskes}},\ and\ \bibinfo {author} {\bibfnamefont {Y.}~\bibnamefont
  {Mishin}},\ }\bibfield  {title} {\bibinfo {title} {Angular-dependent
  interatomic potential for the {Cu}--{Ta} system and its application to
  structural stability of nano-crystalline alloys},\ }\href
  {https://doi.org/10.1016/j.actamat.2015.08.052} {\bibfield  {journal}
  {\bibinfo  {journal} {Acta Materialia}\ }\textbf {\bibinfo {volume} {100}},\
  \bibinfo {pages} {377} (\bibinfo {year} {2015})}\BibitemShut {NoStop}%
\bibitem [{\citenamefont {Starikov}\ \emph {et~al.}(2021)\citenamefont
  {Starikov}, \citenamefont {Smirnova}, \citenamefont {Pradhan}, \citenamefont
  {Lysogorskiy}, \citenamefont {Chapman}, \citenamefont {Mrovec},\ and\
  \citenamefont {Drautz}}]{starikov_2021}%
  \BibitemOpen
  \bibfield  {author} {\bibinfo {author} {\bibfnamefont {S.}~\bibnamefont
  {Starikov}}, \bibinfo {author} {\bibfnamefont {D.}~\bibnamefont {Smirnova}},
  \bibinfo {author} {\bibfnamefont {T.}~\bibnamefont {Pradhan}}, \bibinfo
  {author} {\bibfnamefont {Y.}~\bibnamefont {Lysogorskiy}}, \bibinfo {author}
  {\bibfnamefont {H.}~\bibnamefont {Chapman}}, \bibinfo {author} {\bibfnamefont
  {M.}~\bibnamefont {Mrovec}},\ and\ \bibinfo {author} {\bibfnamefont
  {R.}~\bibnamefont {Drautz}},\ }\bibfield  {title} {\bibinfo {title}
  {Angular-dependent interatomic potential for large-scale atomistic simulation
  of iron: {Development} and comprehensive comparison with existing interatomic
  models},\ }\href {https://doi.org/10.1103/PhysRevMaterials.5.063607}
  {\bibfield  {journal} {\bibinfo  {journal} {Physical Review Materials}\
  }\textbf {\bibinfo {volume} {5}},\ \bibinfo {pages} {063607} (\bibinfo {year}
  {2021})}\BibitemShut {NoStop}%
\bibitem [{\citenamefont {Singh}\ \emph {et~al.}(2020)\citenamefont {Singh},
  \citenamefont {Sharma},\ and\ \citenamefont {Parashar}}]{singh_2020}%
  \BibitemOpen
  \bibfield  {author} {\bibinfo {author} {\bibfnamefont {D.}~\bibnamefont
  {Singh}}, \bibinfo {author} {\bibfnamefont {P.}~\bibnamefont {Sharma}},\ and\
  \bibinfo {author} {\bibfnamefont {A.}~\bibnamefont {Parashar}},\ }\bibfield
  {title} {\bibinfo {title} {Atomistic simulations to study point defect
  dynamics in bi-crystalline niobium},\ }\href
  {https://doi.org/10.1016/j.matchemphys.2020.123628} {\bibfield  {journal}
  {\bibinfo  {journal} {Materials Chemistry and Physics}\ }\textbf {\bibinfo
  {volume} {255}},\ \bibinfo {pages} {123628} (\bibinfo {year}
  {2020})}\BibitemShut {NoStop}%
\bibitem [{\citenamefont {Hahn}\ \emph {et~al.}(2016)\citenamefont {Hahn},
  \citenamefont {Fensin}, \citenamefont {Germann},\ and\ \citenamefont
  {Meyers}}]{hahn_2016}%
  \BibitemOpen
  \bibfield  {author} {\bibinfo {author} {\bibfnamefont {E.~N.}\ \bibnamefont
  {Hahn}}, \bibinfo {author} {\bibfnamefont {S.~J.}\ \bibnamefont {Fensin}},
  \bibinfo {author} {\bibfnamefont {T.~C.}\ \bibnamefont {Germann}},\ and\
  \bibinfo {author} {\bibfnamefont {M.~A.}\ \bibnamefont {Meyers}},\ }\bibfield
   {title} {\bibinfo {title} {Symmetric tilt boundaries in body-centered cubic
  tantalum},\ }\href {https://doi.org/10.1016/j.scriptamat.2016.01.038}
  {\bibfield  {journal} {\bibinfo  {journal} {Scripta Materialia}\ }\textbf
  {\bibinfo {volume} {116}},\ \bibinfo {pages} {108} (\bibinfo {year}
  {2016})}\BibitemShut {NoStop}%
\bibitem [{\citenamefont {Waters}\ \emph {et~al.}(2023)\citenamefont {Waters},
  \citenamefont {Karls}, \citenamefont {Nikiforov}, \citenamefont {Elliott},
  \citenamefont {Tadmor},\ and\ \citenamefont {Runnels}}]{waters_2023}%
  \BibitemOpen
  \bibfield  {author} {\bibinfo {author} {\bibfnamefont {B.}~\bibnamefont
  {Waters}}, \bibinfo {author} {\bibfnamefont {D.~S.}\ \bibnamefont {Karls}},
  \bibinfo {author} {\bibfnamefont {I.}~\bibnamefont {Nikiforov}}, \bibinfo
  {author} {\bibfnamefont {R.~S.}\ \bibnamefont {Elliott}}, \bibinfo {author}
  {\bibfnamefont {E.~B.}\ \bibnamefont {Tadmor}},\ and\ \bibinfo {author}
  {\bibfnamefont {B.}~\bibnamefont {Runnels}},\ }\bibfield  {title} {\bibinfo
  {title} {Automated determination of grain boundary energy and
  potential-dependence using the {OpenKIM} framework},\ }\href
  {https://doi.org/10.1016/j.commatsci.2023.112057} {\bibfield  {journal}
  {\bibinfo  {journal} {Computational Materials Science}\ }\textbf {\bibinfo
  {volume} {220}},\ \bibinfo {pages} {112057} (\bibinfo {year}
  {2023})}\BibitemShut {NoStop}%
\bibitem [{\citenamefont {Frolov}\ \emph
  {et~al.}(2018{\natexlab{b}})\citenamefont {Frolov}, \citenamefont {Zhu},
  \citenamefont {Oppelstrup}, \citenamefont {Marian},\ and\ \citenamefont
  {Rudd}}]{frolov_2018_W}%
  \BibitemOpen
  \bibfield  {author} {\bibinfo {author} {\bibfnamefont {T.}~\bibnamefont
  {Frolov}}, \bibinfo {author} {\bibfnamefont {Q.}~\bibnamefont {Zhu}},
  \bibinfo {author} {\bibfnamefont {T.}~\bibnamefont {Oppelstrup}}, \bibinfo
  {author} {\bibfnamefont {J.}~\bibnamefont {Marian}},\ and\ \bibinfo {author}
  {\bibfnamefont {R.~E.}\ \bibnamefont {Rudd}},\ }\bibfield  {title} {\bibinfo
  {title} {Structures and transitions in bcc tungsten grain boundaries and
  their role in the absorption of point defects},\ }\href
  {https://doi.org/10.1016/j.actamat.2018.07.051} {\bibfield  {journal}
  {\bibinfo  {journal} {Acta Materialia}\ }\textbf {\bibinfo {volume} {159}},\
  \bibinfo {pages} {123} (\bibinfo {year} {2018}{\natexlab{b}})}\BibitemShut
  {NoStop}%
\bibitem [{\citenamefont {Wang}\ \emph {et~al.}(2018)\citenamefont {Wang},
  \citenamefont {Madsen},\ and\ \citenamefont {Drautz}}]{wang_2018}%
  \BibitemOpen
  \bibfield  {author} {\bibinfo {author} {\bibfnamefont {J.}~\bibnamefont
  {Wang}}, \bibinfo {author} {\bibfnamefont {G.~K.~H.}\ \bibnamefont
  {Madsen}},\ and\ \bibinfo {author} {\bibfnamefont {R.}~\bibnamefont
  {Drautz}},\ }\bibfield  {title} {\bibinfo {title} {Grain boundaries in
  bcc-{Fe}: A density-functional theory and tight-binding study},\ }\href
  {https://doi.org/10.1088/1361-651X/aa9f81} {\bibfield  {journal} {\bibinfo
  {journal} {Modelling and Simulation in Materials Science and Engineering}\
  }\textbf {\bibinfo {volume} {26}},\ \bibinfo {pages} {025008} (\bibinfo
  {year} {2018})}\BibitemShut {NoStop}%
\bibitem [{\citenamefont {Larsen}\ \emph {et~al.}(2017)\citenamefont {Larsen},
  \citenamefont {Mortensen}, \citenamefont {Blomqvist}, \citenamefont
  {Castelli}, \citenamefont {Christensen}, \citenamefont {Dulak}, \citenamefont
  {Friis}, \citenamefont {Groves}, \citenamefont {Hammer}, \citenamefont
  {Hargus} \emph {et~al.}}]{larsen_2017}%
  \BibitemOpen
  \bibfield  {author} {\bibinfo {author} {\bibfnamefont {A.~H.}\ \bibnamefont
  {Larsen}}, \bibinfo {author} {\bibfnamefont {J.~J.}\ \bibnamefont
  {Mortensen}}, \bibinfo {author} {\bibfnamefont {J.}~\bibnamefont
  {Blomqvist}}, \bibinfo {author} {\bibfnamefont {I.~E.}\ \bibnamefont
  {Castelli}}, \bibinfo {author} {\bibfnamefont {R.}~\bibnamefont
  {Christensen}}, \bibinfo {author} {\bibfnamefont {M.}~\bibnamefont {Dulak}},
  \bibinfo {author} {\bibfnamefont {J.}~\bibnamefont {Friis}}, \bibinfo
  {author} {\bibfnamefont {M.~N.}\ \bibnamefont {Groves}}, \bibinfo {author}
  {\bibfnamefont {B.}~\bibnamefont {Hammer}}, \bibinfo {author} {\bibfnamefont
  {C.}~\bibnamefont {Hargus}}, \emph {et~al.},\ }\bibfield  {title} {\bibinfo
  {title} {The atomic simulation environment---a {Python} library for working
  with atoms},\ }\href {https://doi.org/10.1088/1361-648X/aa680e} {\bibfield
  {journal} {\bibinfo  {journal} {Journal of Physics: Condensed Matter}\
  }\textbf {\bibinfo {volume} {29}},\ \bibinfo {pages} {273002} (\bibinfo
  {year} {2017})}\BibitemShut {NoStop}%
\bibitem [{\citenamefont {Thompson}\ \emph {et~al.}(2022)\citenamefont
  {Thompson}, \citenamefont {Aktulga}, \citenamefont {Berger}, \citenamefont
  {Bolintineanu}, \citenamefont {Brown}, \citenamefont {Crozier}, \citenamefont
  {{in 't Veld}}, \citenamefont {Kohlmeyer}, \citenamefont {Moore},
  \citenamefont {Nguyen} \emph {et~al.}}]{thompson_2022}%
  \BibitemOpen
  \bibfield  {author} {\bibinfo {author} {\bibfnamefont {A.~P.}\ \bibnamefont
  {Thompson}}, \bibinfo {author} {\bibfnamefont {H.~M.}\ \bibnamefont
  {Aktulga}}, \bibinfo {author} {\bibfnamefont {R.}~\bibnamefont {Berger}},
  \bibinfo {author} {\bibfnamefont {D.~S.}\ \bibnamefont {Bolintineanu}},
  \bibinfo {author} {\bibfnamefont {W.~M.}\ \bibnamefont {Brown}}, \bibinfo
  {author} {\bibfnamefont {P.~S.}\ \bibnamefont {Crozier}}, \bibinfo {author}
  {\bibfnamefont {P.~J.}\ \bibnamefont {{in 't Veld}}}, \bibinfo {author}
  {\bibfnamefont {A.}~\bibnamefont {Kohlmeyer}}, \bibinfo {author}
  {\bibfnamefont {S.~G.}\ \bibnamefont {Moore}}, \bibinfo {author}
  {\bibfnamefont {T.~D.}\ \bibnamefont {Nguyen}}, \emph {et~al.},\ }\bibfield
  {title} {\bibinfo {title} {{LAMMPS} - a flexible simulation tool for
  particle-based materials modeling at the atomic, meso, and continuum
  scales},\ }\href {https://doi.org/10.1016/j.cpc.2021.108171} {\bibfield
  {journal} {\bibinfo  {journal} {Computer Physics Communications}\ }\textbf
  {\bibinfo {volume} {271}},\ \bibinfo {pages} {108171} (\bibinfo {year}
  {2022})}\BibitemShut {NoStop}%
\bibitem [{\citenamefont {Kresse}\ and\ \citenamefont
  {Hafner}(1993)}]{kresse_1993}%
  \BibitemOpen
  \bibfield  {author} {\bibinfo {author} {\bibfnamefont {G.}~\bibnamefont
  {Kresse}}\ and\ \bibinfo {author} {\bibfnamefont {J.}~\bibnamefont
  {Hafner}},\ }\bibfield  {title} {\bibinfo {title} {\emph{Ab initio} molecular
  dynamics for liquid metals},\ }\href
  {https://doi.org/10.1103/PhysRevB.47.558} {\bibfield  {journal} {\bibinfo
  {journal} {Physical Review B}\ }\textbf {\bibinfo {volume} {47}},\ \bibinfo
  {pages} {558} (\bibinfo {year} {1993})}\BibitemShut {NoStop}%
\bibitem [{\citenamefont {Kresse}\ and\ \citenamefont
  {Hafner}(1994)}]{kresse_1994}%
  \BibitemOpen
  \bibfield  {author} {\bibinfo {author} {\bibfnamefont {G.}~\bibnamefont
  {Kresse}}\ and\ \bibinfo {author} {\bibfnamefont {J.}~\bibnamefont
  {Hafner}},\ }\bibfield  {title} {\bibinfo {title} {\emph{Ab initio}
  molecular-dynamics simulation of the liquid-metal--amorphous-semiconductor
  transition in germanium},\ }\href {https://doi.org/10.1103/PhysRevB.49.14251}
  {\bibfield  {journal} {\bibinfo  {journal} {Physical Review B}\ }\textbf
  {\bibinfo {volume} {49}},\ \bibinfo {pages} {14251} (\bibinfo {year}
  {1994})}\BibitemShut {NoStop}%
\bibitem [{\citenamefont {Kresse}\ and\ \citenamefont
  {Furthm{\"u}ller}(1996{\natexlab{a}})}]{kresse_1996a}%
  \BibitemOpen
  \bibfield  {author} {\bibinfo {author} {\bibfnamefont {G.}~\bibnamefont
  {Kresse}}\ and\ \bibinfo {author} {\bibfnamefont {J.}~\bibnamefont
  {Furthm{\"u}ller}},\ }\bibfield  {title} {\bibinfo {title} {Efficient
  iterative schemes for \emph{ab initio} total-energy calculations using a
  plane-wave basis set},\ }\href {https://doi.org/10.1103/PhysRevB.54.11169}
  {\bibfield  {journal} {\bibinfo  {journal} {Physical Review B}\ }\textbf
  {\bibinfo {volume} {54}},\ \bibinfo {pages} {11169} (\bibinfo {year}
  {1996}{\natexlab{a}})}\BibitemShut {NoStop}%
\bibitem [{\citenamefont {Kresse}\ and\ \citenamefont
  {Furthm{\"u}ller}(1996{\natexlab{b}})}]{kresse_1996b}%
  \BibitemOpen
  \bibfield  {author} {\bibinfo {author} {\bibfnamefont {G.}~\bibnamefont
  {Kresse}}\ and\ \bibinfo {author} {\bibfnamefont {J.}~\bibnamefont
  {Furthm{\"u}ller}},\ }\bibfield  {title} {\bibinfo {title} {Efficiency of
  \emph{ab initio} total energy calculations for metals and semiconductors
  using a plane-wave basis set},\ }\href
  {https://doi.org/10.1016/0927-0256(96)00008-0} {\bibfield  {journal}
  {\bibinfo  {journal} {Computational Materials Science}\ }\textbf {\bibinfo
  {volume} {6}},\ \bibinfo {pages} {15} (\bibinfo {year}
  {1996}{\natexlab{b}})}\BibitemShut {NoStop}%
\bibitem [{\citenamefont {Kresse}\ and\ \citenamefont
  {Joubert}(1999)}]{kresse_1999}%
  \BibitemOpen
  \bibfield  {author} {\bibinfo {author} {\bibfnamefont {G.}~\bibnamefont
  {Kresse}}\ and\ \bibinfo {author} {\bibfnamefont {D.}~\bibnamefont
  {Joubert}},\ }\bibfield  {title} {\bibinfo {title} {From ultrasoft
  pseudopotentials to the projector augmented-wave method},\ }\href
  {https://doi.org/10.1103/PhysRevB.59.1758} {\bibfield  {journal} {\bibinfo
  {journal} {Physical Review B}\ }\textbf {\bibinfo {volume} {59}},\ \bibinfo
  {pages} {1758} (\bibinfo {year} {1999})}\BibitemShut {NoStop}%
\bibitem [{\citenamefont {Perdew}\ \emph {et~al.}(1996)\citenamefont {Perdew},
  \citenamefont {Burke},\ and\ \citenamefont {Ernzerhof}}]{perdew_1996}%
  \BibitemOpen
  \bibfield  {author} {\bibinfo {author} {\bibfnamefont {J.~P.}\ \bibnamefont
  {Perdew}}, \bibinfo {author} {\bibfnamefont {K.}~\bibnamefont {Burke}},\ and\
  \bibinfo {author} {\bibfnamefont {M.}~\bibnamefont {Ernzerhof}},\ }\bibfield
  {title} {\bibinfo {title} {Generalized gradient approximation made simple},\
  }\href {https://doi.org/10.1103/PhysRevLett.77.3865} {\bibfield  {journal}
  {\bibinfo  {journal} {Physical Review Letters}\ }\textbf {\bibinfo {volume}
  {77}},\ \bibinfo {pages} {3865} (\bibinfo {year} {1996})}\BibitemShut
  {NoStop}%
\bibitem [{\citenamefont {Monkhorst}\ and\ \citenamefont
  {Pack}(1976)}]{monkhorst_1976}%
  \BibitemOpen
  \bibfield  {author} {\bibinfo {author} {\bibfnamefont {H.~J.}\ \bibnamefont
  {Monkhorst}}\ and\ \bibinfo {author} {\bibfnamefont {J.~D.}\ \bibnamefont
  {Pack}},\ }\bibfield  {title} {\bibinfo {title} {Special points for
  {Brillouin}-zone integrations},\ }\href
  {https://doi.org/10.1103/PhysRevB.13.5188} {\bibfield  {journal} {\bibinfo
  {journal} {Physical Review B}\ }\textbf {\bibinfo {volume} {13}},\ \bibinfo
  {pages} {5188} (\bibinfo {year} {1976})}\BibitemShut {NoStop}%
\bibitem [{\citenamefont {Ong}\ \emph {et~al.}(2013)\citenamefont {Ong},
  \citenamefont {Richards}, \citenamefont {Jain}, \citenamefont {Hautier},
  \citenamefont {Kocher}, \citenamefont {Cholia}, \citenamefont {Gunter},
  \citenamefont {Chevrier}, \citenamefont {Persson},\ and\ \citenamefont
  {Ceder}}]{ong_2013}%
  \BibitemOpen
  \bibfield  {author} {\bibinfo {author} {\bibfnamefont {S.~P.}\ \bibnamefont
  {Ong}}, \bibinfo {author} {\bibfnamefont {W.~D.}\ \bibnamefont {Richards}},
  \bibinfo {author} {\bibfnamefont {A.}~\bibnamefont {Jain}}, \bibinfo {author}
  {\bibfnamefont {G.}~\bibnamefont {Hautier}}, \bibinfo {author} {\bibfnamefont
  {M.}~\bibnamefont {Kocher}}, \bibinfo {author} {\bibfnamefont
  {S.}~\bibnamefont {Cholia}}, \bibinfo {author} {\bibfnamefont
  {D.}~\bibnamefont {Gunter}}, \bibinfo {author} {\bibfnamefont {V.~L.}\
  \bibnamefont {Chevrier}}, \bibinfo {author} {\bibfnamefont {K.~A.}\
  \bibnamefont {Persson}},\ and\ \bibinfo {author} {\bibfnamefont
  {G.}~\bibnamefont {Ceder}},\ }\bibfield  {title} {\bibinfo {title} {{Python}
  {Materials} {Genomics} ({pymatgen}): A robust, open-source {Python} library
  for materials analysis},\ }\href
  {https://doi.org/10.1016/j.commatsci.2012.10.028} {\bibfield  {journal}
  {\bibinfo  {journal} {Computational Materials Science}\ }\textbf {\bibinfo
  {volume} {68}},\ \bibinfo {pages} {314} (\bibinfo {year} {2013})}\BibitemShut
  {NoStop}%
\bibitem [{\citenamefont {Methfessel}\ and\ \citenamefont
  {Paxton}(1989)}]{methfessel_1989}%
  \BibitemOpen
  \bibfield  {author} {\bibinfo {author} {\bibfnamefont {M.}~\bibnamefont
  {Methfessel}}\ and\ \bibinfo {author} {\bibfnamefont {A.~T.}\ \bibnamefont
  {Paxton}},\ }\bibfield  {title} {\bibinfo {title} {High-precision sampling
  for {Brillouin}-zone integration in metals},\ }\href
  {https://doi.org/10.1103/PhysRevB.40.3616} {\bibfield  {journal} {\bibinfo
  {journal} {Physical Review B}\ }\textbf {\bibinfo {volume} {40}},\ \bibinfo
  {pages} {3616} (\bibinfo {year} {1989})}\BibitemShut {NoStop}%
\bibitem [{\citenamefont {Loane}\ \emph {et~al.}(1991)\citenamefont {Loane},
  \citenamefont {Xu},\ and\ \citenamefont {Silcox}}]{loane_1991}%
  \BibitemOpen
  \bibfield  {author} {\bibinfo {author} {\bibfnamefont {R.~F.}\ \bibnamefont
  {Loane}}, \bibinfo {author} {\bibfnamefont {P.}~\bibnamefont {Xu}},\ and\
  \bibinfo {author} {\bibfnamefont {J.}~\bibnamefont {Silcox}},\ }\bibfield
  {title} {\bibinfo {title} {Thermal vibrations in convergent-beam electron
  diffraction},\ }\href {https://doi.org/10.1107/S0108767391000375} {\bibfield
  {journal} {\bibinfo  {journal} {Acta Crystallographica Section A}\ }\textbf
  {\bibinfo {volume} {47}},\ \bibinfo {pages} {267} (\bibinfo {year}
  {1991})}\BibitemShut {NoStop}%
\bibitem [{\citenamefont {Kirkland}(2020)}]{kirkland_2020}%
  \BibitemOpen
  \bibfield  {author} {\bibinfo {author} {\bibfnamefont {E.~J.}\ \bibnamefont
  {Kirkland}},\ }\href {https://doi.org/10.1007/978-3-030-33260-0} {\emph
  {\bibinfo {title} {Advanced Computing in Electron Microscopy}}}\ (\bibinfo
  {publisher} {Springer International Publishing},\ \bibinfo {address} {Cham},\
  \bibinfo {year} {2020})\BibitemShut {NoStop}%
\end{thebibliography}%
